\documentclass[preprint]{ptephy_v2}

\preprintnumber{XXXX-XXXX} 
\usepackage{bm}
\usepackage{amsmath}
\usepackage{graphicx}
\usepackage{url}
\usepackage{dcolumn}
\usepackage[colorlinks=true, linkcolor=blue, citecolor=blue, urlcolor=blue]{hyperref}
\usepackage{subcaption}
\usepackage{float}
\usepackage{braket}

\begin{document}

\title{Shell-model calculation with density-dependent interaction for $pf$-shell nuclei}

\author[1]{Kota Yoshinaga}

\author[2]{Noritaka Shimizu} 
\author[2,3]{Takashi Nakatsukasa}

\affil[1]{Graduate School of Science and Technology, University of Tsukuba, Tsukuba 305-8571, Japan}

\affil[2]{Center for Computational Sciences and 
Institute of Pure and Applied Sciences, University of Tsukuba, Tsukuba 305-8577, Japan}

\affil[3]{RIKEN Nishina Center, Wako 351-0198, Japan}

\begin{abstract}
Shell-model calculations with density-dependent interactions are 
performed to investigate $pf$-shell nuclei,
examining the ground-state energies,
low-lying spectra, and $E2$ transition probabilities.
The density-dependent terms in the interaction are self-consistently determined
using the shell-model wave function for the ground state.
We test three density-dependent interactions
adapted from density functionals of Gogny-D1S, Gogny-GT2, and M3Y-P6.
The shell-model results satisfactorily agree with the experimental data.
However, the Gogny-D1S and Gogny-GT2 fail to reproduce the magicity of $N=28$,
while it is properly described by the M3Y-P6 functional.
\end{abstract}

\subjectindex{D10,D11}

\maketitle

\section{\label{sec:intro}Introduction}

In studies of nuclear structure, configuration mixing is often essential to describe low-lying spectra of atomic nuclei. The nuclear shell-model (SM) calculation is a powerful method for investigating various exotic phenomena, \textit{e.g.}, shape coexistence \cite{RevModPhys.83.1467} and the emergence of new magic numbers accompanied by shell evolution \cite{otsuka2020evolution}. This method fully incorporates the effects of configuration mixing in a restricted model space. However, for a high-precision description, the SM interaction must be tuned phenomenologically for each mass region and its corresponding model space. This procedure is a challenging task, especially for nuclei in heavy-mass regions. Although \textit{ab initio} derivations of the SM Hamiltonian have recently achieved great progress and have reached a precision comparable to that of the phenomenological Hamiltonian, it is still difficult to apply it to heavy-mass regions where the intruder orbits play a major role \cite{stroberg2019nonempirical}. 

In contrast to the SM, the density functional theory (DFT) covers the whole mass region of the nuclear chart with a single universal density functional \cite{RevModPhys.75.121}.
Using a large single-particle space in the mean-field level,
both the stationary and the time-dependent DFT calculations have been successful in the description of various properties of nuclear structure, responses, and reactions \cite{RevModPhys.88.045004}.
The configuration-mixing calculations beyond the mean field can be performed, in principle, 
\textit{e.g.}, using the generator coordinate method (GCM) \cite{HillWheeler,GriffinWheeler,RingSchuck1980}.
However, naive applications of the GCM and fully variational approaches with the configuration mixing often lead to unphysical solutions
\cite{VAPDFT,StochasticGCM1,StochasticGCM2,RevModPhys.88.045004}.
In addition, the fractional power of the density in the energy density functional (EDF) often causes a serious problem with the configuration mixing \cite{fractionalpower_PhysRevC.76.054315}.
It is desirable to establish a configuration mixing method based on the DFT with high reliability and numerical stability.

Various efforts have been made to construct a hybrid approach by combining the two contrasting theoretical models, SM and DFT.
A possible approach for such a direction is to prepare the SM two-body matrix elements (TBME) using a density-dependent interaction for the EDF.
The density-dependent interaction is determined by using a one-body density of a certain state, for instance, of the Hartree-Fock solution.
Pioneering works of the SM calculations employing the Skyrme functional were performed for some nearly closed-shell nuclei in Ref.~\cite{SAGAWA1985228} and the $p$-shell nuclei in Ref.~\cite{GOMEZ1993451}.
However, there is a difficulty in applying the method with Skyrme interactions to heavy nuclei.
For nuclei in the medium-heavy-mass regions, inclusion of the pairing EDF is indispensable for their quantitative descriptions.
For the Skyrme DFT, it is customary to add the pairing EDF in the particle-particle (pairing) channel to the Skyrme EDF in the particle-hole channel.
The pairing EDF is usually associated with
a zero-range density-dependent interaction
different from the Skyrme interaction\footnote{
There are some exceptions, such as SkP \cite{DOBACZEWSKI1984103},
in which the same Skyrme interaction is adopted
in both the particle-hole and the particle-particle channels,
however, its isovector components are not well constrained \cite{PhysRevC.86.034333}.
}.
Therefore, it is not trivial to apply the Skyrme interactions to the calculation of TBME's for the SM calculation.
Note that, in parallel, several attempts have been also made to combine covariant DFT with SM configuration mixing \cite{Liu_2025, PhysRevLett.132.232501}.

In contrast to the Skyrme EDF,
the Gogny EDF, which consists of two-body interactions of the finite-range Gaussian form, is designed to give proper correlations in both particle-particle and particle-hole channels \cite{PhysRevC.21.1568}. Thus, it is suitable for the calculation of the TBMEs in the SM interaction. 
Reference~\cite{PhysRevC.98.044320} demonstrated that the SM Hamiltonian constructed by the Gogny-D1S EDF provides a good description of $p$- and $sd$-shell nuclei. We also reported further $sd$-shell results in Ref.~\cite{particles8020061}.
References~\cite{PhysRevC.89.011306,PhysRevC.95.044315} proposed a variational method to determine the single-particle wave function with configuration mixing using the Gogny-D1S functional, and discussed its feasibility for $sd$-shell nuclei thoroughly.
However, it is still unclear whether such a hybrid method can be applicable to heavier nuclei, such as $pf$-shell nuclei and beyond.

In the present article, we perform SM calculations utilizing finite-range density-dependent interactions for $pf$-shell nuclei in the $0\hbar\omega$ model space and discuss their performance by comparing the results with the experimental data and with conventional SM calculations.
The density is determined self-consistently by employing the SM ground-state wave function following the prescription given in Ref.~\cite{PhysRevC.98.044320}.
We employ three EDFs: Gogny-D1S \cite{PhysRevC.21.1568}, Gogny-GT2 \cite{PhysRevLett.97.162501}, and M3Y-P6 \cite{PhysRevC.87.014336}. The Gogny-D1S is the most widely used Gogny EDF and has proved its usefulness in Ref.~\cite{PhysRevC.98.044320}, however, the tensor force is not explicitly included in the functional. The Gogny-GT2 parametrization was proposed in Ref.~\cite {PhysRevLett.97.162501}, including the tensor force explicitly to reproduce the evolution of shell structures in unstable nuclei, which could be advantageous for the current SM study. The M3Y-P6 functional consists of Yukawa-type functions, instead of Gaussian functions in the Gogny EDFs. The parameters of M3Y-P6 are fitted to reproduce the $G$-matrix interaction. We will present the results for the $pf$-shell nuclei using these functionals, together with the experimental data and the conventional SM results given by the GXPF1A interaction \cite{Honma2005}. 

The paper is organized as follows:
In Sect.~\ref{sec:theor}, we describe the present theoretical framework of the DFT-SM hybrid model.
Section~\ref{chap:result} is devoted to the numerical results.
We discuss the TBMEs constructed from the density-dependent interactions in Sect.~\ref{sec:TBME}.
The SM ground-state energies and ground-state spins are presented in Sect.~\ref{sec:gs}. 
The excitation spectra and $E2$ transition probabilities are shown in Sect.~\ref{sec:excited}, and finally, the magicity of $^{56}$Ni is examined in Sect.~\ref{sec:magicity}.
The summary is given in Sect.~\ref{sec:summary}.

\section{\label{sec:theor}Theoretical framework}

\subsection{\label{sec:smfram}Shell-model Hamiltonian}

We present a method to construct the SM Hamiltonian employing density-dependent interaction adapted from density functionals.
First, we construct the Hamiltonian, $H_\textrm{HO}$, which contains one-body and two-body terms
whose matrix elements are calculated in the harmonic-oscillator (HO) basis:
\begin{equation}
    H_\textrm{HO} = \sum_{i,j} {T_{ij}c_{i}^{\dagger}c_{j}} + \frac14 \sum\limits_{i, j, k, l} {v[\rho]_{ijkl}c_{i}^{\dagger}c_{j}^{\dagger}c_{l}c_{k}}
  \label{eq:nocoreH},
\end{equation}
where $c^\dagger_i$ denotes the creation operator of a nucleon at the state $i$ of the HO basis.
The frequency $\omega$ of the HO potential is 
taken from an empirical formula $\hbar\omega=45A^{-1/3}-25A^{-2/3}$, and its validity is discussed in Sect.~\ref{chap:result}.
The one-body part is given by the kinetic term $T_{ij}=\langle i|\frac{\bf{p}^2}{2m} | j \rangle$,
and the antisymmetrized TBMEs
are evaluated with the density-dependent interaction at a given density $\rho(\mathbf{r})$ as
\begin{equation}
v(\rho)_{ijkl}=\braket{ij|V[\rho]|kl}-\braket{ij|V[\rho]|lk} ,
\end{equation}
where the two-body potential $V[\rho]$ is given in Sect.~\ref{sec:ddint}.
The TBMEs in the HO basis are efficiently calculated employing the Fourier transformation \cite{horie1961energy}.
We neglect a violation of the isospin symmetry for simplicity, thus, the isospin symmetry is conserved in the present study.
The Coulomb energy is empirically added for the calculation of the total energy.

The Hamiltonian $H_\textrm{HO}$ of Eq.~(\ref{eq:nocoreH}) is defined in a no-core space.
In the present work, we assume $^{40}$Ca as an inert core and $pf$ shell is taken as the model space,
namely the $0f_{7/2}, 0f_{5/2}, 1p_{3/2},$ and $1p_{1/2}$ orbits. 
We obtain the SM Hamiltonian of the $pf$-shell model space as
\begin{equation}
    H_\textrm{SM} = \sum_{i} {\epsilon_{i}[\rho] c_{i}^{\dagger}c_{i}} + \frac14 \sum\limits_{i, j, k, l} {v[\rho]_{ijkl}c_{i}^{\dagger}c_{j}^{\dagger}c_{l}c_{k}}
  \label{eq:smH},
\end{equation}
where the $i,j,k,$ and $l$ indices denote the single-particle states inside the model space.
The single-particle energies are given by the sum of the kinetic and the potential terms from the inert core as
\begin{equation}
    \epsilon_{i}[\rho] = T_{ii} + \sum_n v[\rho]_{inin} , 
\end{equation}
where $n$ runs over the occupied states of the inert core ($^{40}$Ca)
and $i$ is a single-particle state in the $pf$ shell.
Non-diagonal one-body matrix elements do not appear because of the $0\hbar\omega$ model space.
This Hamiltonian is rewritten in the $J$-coupled form as
\begin{equation}
    H_\textrm{SM} = \sum_a \epsilon_a[\rho] n_a + \sum_{a\leq b, c \leq d, J, M, T, T_z} \langle a,b,|v[\rho]|c,d\rangle_{JT}\  A^\dagger(a,b,J,M,T,T_z) A(c,d,J,M,T,T_z),
    \label{eq:H_sm}
\end{equation}
where $n_a$ denotes the number operator of the single-particle orbit $a$, and $A^\dagger(a,b,J,M,T,T_z)$ is the creation operator of a nucleon pair coupled to the angular-momentum $(J,M)$ and isospin $(T,T_z)$.
$\epsilon_a[\rho]$ and $\langle a,b,|v[\rho]|c,d\rangle_{JT}$ are the single-particle energy and the TBME, respectively, which depend on the density $\rho(\mathbf{r})$.

Performing the SM calculation using the Hamiltonian of Eq.~(\ref{eq:H_sm}),
we obtain the SM wave function for the ground state $\ket{0}$ and its ground-state density
$\rho(\mathbf{r})\equiv\sum_{\sigma\tau}\braket{0 | \psi^\dagger_{\sigma\tau}(\mathbf{r})\psi_{\sigma\tau}(\mathbf{r}) | 0}$.
Updating $\rho(\mathbf{r})$, we reconstruct the SM Hamiltonian and perform the SM calculation iteratively till convergence,
following the procedure proposed in Ref.~\cite{PhysRevC.98.044320}. 
Typically, three iterations are required for even-even-mass nuclei and ten iterations for the other nuclei.
Note that the local density of the ground state is used to construct the TBMEs.
Therefore, the fractional power of density does not cause a problem.
For certain odd-mass and odd-odd nuclei, the ground-state spin cannot be specified in advance. In such a case, we perform the iterative calculations for all possible spins and adopt the density corresponding to the lowest energy.

After completing these iterations, the ground-state energy is evaluated as 
\begin{equation}
 E =  E_\textrm{SM} + E_\textrm{core} + E_\textrm{Coul} -T_\textrm{CoM},  
 \label{eq:Egs}
\end{equation}
where $E_\textrm{core}$ is the expectation value of $^{40}$Ca with the Hamiltonian in Eq.~(\ref{eq:nocoreH}).
$E_\textrm{SM}$ is the eigenvalue of Eq.~(\ref{eq:H_sm}),  and $T_\textrm{CoM}=\frac34\hbar\omega$ is the kinetic energy of the center-of-mass motion \cite{PhysRevC.98.044320}. $E_\textrm{Coul}$ is the contribution of the Coulomb force and is estimated using an empirical formula as  
 \cite{PhysRevC.59.2033,PhysRevC.97.054321}
\begin{equation}
    E_{\rm{Coul.}} = 0.700 \frac{Z(Z-1) - 0.76 (Z(Z-1))^{\frac{2}{3}}}{e^{\frac{3}{2A}}A^{\frac{1}{3}}(0.946 - 0.573 (\frac{|Z-N|}{A})^{2})} \ \mathrm{MeV}, \notag
\end{equation}
where $e$ denotes the base of the natural logarithm. This energy does not affect excitation energies.
Because of the isospin symmetry of the Hamiltonian,
the excitation spectra in mirror nuclei are identical to each other.

\subsection{\label{sec:ddint}Density-dependent interactions}

In this subsection, we briefly review three EDFs used in the present work: Gogny-D1S, Gogny-GT2, and M3Y-P6. 
All these functionals contain a density-dependent zero-range force to describe nuclear saturation.

The Gogny effective interaction with the D1S parameter set consists of two-range Gaussian central terms, a spin-orbit term, and a density-dependent term.
The one of the GT2 parameter set contains these terms and, in addition, includes a tensor term.
The Gogny potential between two nucleons \cite{PhysRevC.21.1568} is expressed analytically in the coordinate space as 
\begin{eqnarray}
  V_{\mathrm{Gogny}}(\bm{r}_1, \bm{r}_2)
  &=& \sum_{i=1}^{2}
     e^{ -(\bm{r}_1 - \bm{r}_2)^2 / \mu_i^2 }
     \Bigl( 
     t_i^{(\mathrm{SE})} P^{(\mathrm{SE})} +
     t_i^{(\mathrm{TE})} P^{(\mathrm{TE})} +
     t_i^{(\mathrm{SO})} P^{(\mathrm{SO})} +
     t_i^{(\mathrm{TO})} P^{(\mathrm{TO})}
     \Bigr) \notag \\
  & &\quad +\, i W_{\mathrm{LS}}
     (\bm{\sigma}_1 + \bm{\sigma}_2)
     \cdot \bm{k'} \times
     \delta( \bm{r}_1 - \bm{r}_2 ) \bm{k} \notag \\
  & &\quad +\, t_3 (1 + x_0 P_{\sigma})
     \delta( \bm{r}_1 - \bm{r}_2 )
     \left[
       \rho\!\left(
         \frac{\bm{r}_1 + \bm{r}_2}{2}
       \right)
     \right]^{\alpha} \notag \\
  & &\quad +\, V_{\mathrm{TS}}
     e^{ -(\bm{r}_1 - \bm{r}_2)^2 / \mu_{\mathrm{TS}}^2 }
     (\bm{\tau}_1 \cdot \bm{\tau}_2) S_{12},
  \label{eq:gognyint}
\end{eqnarray}
where $\bm{r}_{1}$ and $\bm{r}_{2}$ are the coordinates of the two interacting nucleons, and $\bm{k}$ and $\bm{k'}$ are the relative momenta acting on the right and the left, respectively.
$\bm{\sigma}_1$ and $\bm{\sigma}_2$ are the Pauli spin matrices of nucleons 1 and 2, and $\bm{\tau}_1$ and $\bm{\tau}_2$ are their isospin matrices.
$P^{(\mathrm{SE})}$, $P^{(\mathrm{TE})}$, $P^{(\mathrm{SO})}$, and $P^{(\mathrm{TO})}$ are the projection operators of the singlet-even (SE), triplet-even (TE), singlet-odd (SO), and triplet-odd (TO) channels, respectively.
The quantities $\mu_{i}$, $t_i^{(\mathrm{SE})}$,
$t_i^{(\mathrm{TE})}$, $t_i^{(\mathrm{SO})}$, 
$t_i^{(\mathrm{TO})}$, $W_\textrm{LS}$, $V_{\mathrm{TS}}$, $t_{3}$, $x_{0}$, and $\alpha$ are parameters of the Gogny functionals and are summarized in Table \ref{tab:gognyparams} of Appendix \ref{app:paramlist}.

The density-dependent term depends on the nuclear density, and
the parameters of $t_3$, $x_0$, and $\alpha$ are determined to reproduce the nuclear saturation and incompressibility.
The tensor term does not exist in the Gogny-D1S functional, but was introduced in the GT2 parametrization based on the one-pion exchange potential to describe the gradual change of shell structure in Sb isotopes \cite{PhysRevLett.97.162501}.
The tensor operator $S_{12}$ in Eq.~(\ref{eq:gognyint}) is defined as
\begin{equation}
  S_{12} = \frac{3 [\bm\sigma_{1} \cdot (\bm{r}_{1} - \bm{r}_{2})][\bm\sigma_{2} \cdot (\bm{r}_{1} - \bm{r}_{2})]}{(\bm{r}_{1} - \bm{r}_{2})^{2}} - (\bm\sigma_{1} \cdot \bm\sigma_{2}).\notag
\end{equation}

The Michigan three-range Yukawa (M3Y) interaction is known as a semi-realistic interaction based on the $G$-matrix \cite{anantaraman1983effective}. Recently, M3Y-type interactions with the density-dependent interaction was proposed
for DFT calculations to describe the whole region of the nuclear chart \cite{PhysRevC.68.014316,nakada2014predicting}.
The finite-range Yukawa parts consist of the central, tensor, and spin-orbit terms, while
the density-dependent term is zero-range:
\begin{eqnarray}
&&    V_{\mathrm{M3Y}}(\bm{r}_1, \bm{r}_2)
\nonumber \\
&=& \sum_{i=1}^{3}
   f_i^{(\mathrm{C})}(\bm{r}_1, \bm{r}_2)
   \Bigl(
     t_i^{(\mathrm{SE})} P^{(\mathrm{SE})} +
     t_i^{(\mathrm{TE})} P^{(\mathrm{TE})} +
     t_i^{(\mathrm{SO})} P^{(\mathrm{SO})} +
     t_i^{(\mathrm{TO})} P^{(\mathrm{TO})}
   \Bigr) \nonumber \\
& & + \sum_{i=1}^{2}
   f_i^{(\mathrm{LS})}(\bm{r}_1, \bm{r}_2)
   \Bigl(
     t_i^{(\mathrm{LSE})} P^{(\mathrm{TE})} +
     t_i^{(\mathrm{LSO})} P^{(\mathrm{TO})}
   \Bigr)
   \bm{L}_{12} \cdot (\bm{\sigma}_1 + \bm{\sigma}_2) \nonumber \\
& & + \sum_{i=1}^{2}
   f_i^{(\mathrm{TS})}(\bm{r}_1, \bm{r}_2)
   \Bigl(
     t_i^{(\mathrm{TSE})} P^{(\mathrm{TE})}  +
     t_i^{(\mathrm{TSO})} P^{(\mathrm{TO})}
   \Bigr)
   S_{12} \nonumber \\
& & + \delta(\bm{r}_1 - \bm{r}_2)
   \Bigl[
     t_\rho^{(\mathrm{SE})}
     \left\{ \rho\!\left( \tfrac{\bm{r}_1 + \bm{r}_2}{2} \right) \right\}^{\alpha^{(\mathrm{SE})}}
     P^{(\mathrm{SE})} +
     t_\rho^{(\mathrm{TE})}
     \left\{ \rho\!\left( \tfrac{\bm{r}_1 + \bm{r}_2}{2} \right) \right\}^{\alpha^{(\mathrm{TE})}}
     P^{(\mathrm{TE})}
   \Bigr]
   .
\label{eq:m3yint}
\end{eqnarray}
$f_i^{(X)}(\bm{r}_1, \bm{r}_2)$ is the Yukawa-type function defined as 
\begin{equation}
    f_i^{(X)}(\bm{r}_1, \bm{r}_2) = \frac{e^{- |\bm{r}_1 - \bm{r}_2| / \mu_i^{(X)}}} { |\bm{r}_1 - \bm{r}_2| / \mu_i^{(X)}} \notag
\end{equation}
with the finite-range parameter $\mu^{(X)}_i$.
The index $X$ denotes the type of the nucleon-nucleon interaction, namely the central (C), the spin-orbit (LS), the tensor (TS), and the density-dependent (DD) forces.
$L_{12}$ is the relative angular momentum operator and $S_{12}$ the tensor operator between nucleon 1 and nucleon 2.
The parameters of these coupling constants are summarized in Table \ref{tab:m3yparams} of Appendix \ref{app:paramlist}.
In contrast to the Gogny interaction in Eq.~(\ref{eq:gognyint}), the M3Y interaction incorporates both TE and TO channels in its tensor force.
The finite-range character of all the central, tensor, and spin–orbit terms is a key distinction from the Gogny interactions.

In the present study, the theoretical models based on DFT are applied mainly to $pf$-shell nuclei in the $0 \hbar \omega$ model space.
We investigate the nuclear properties in both the ground state and excited states and compare the results of the three theoretical models with the GXPF1A results and experimental data.
The KSHELL code \cite{SHIMIZU2019372} is used for the SM calculations.

\section{Results and discussion}
\label{chap:result}

\begin{figure}[tbp]
  \centering
  \includegraphics[width=0.5\linewidth]{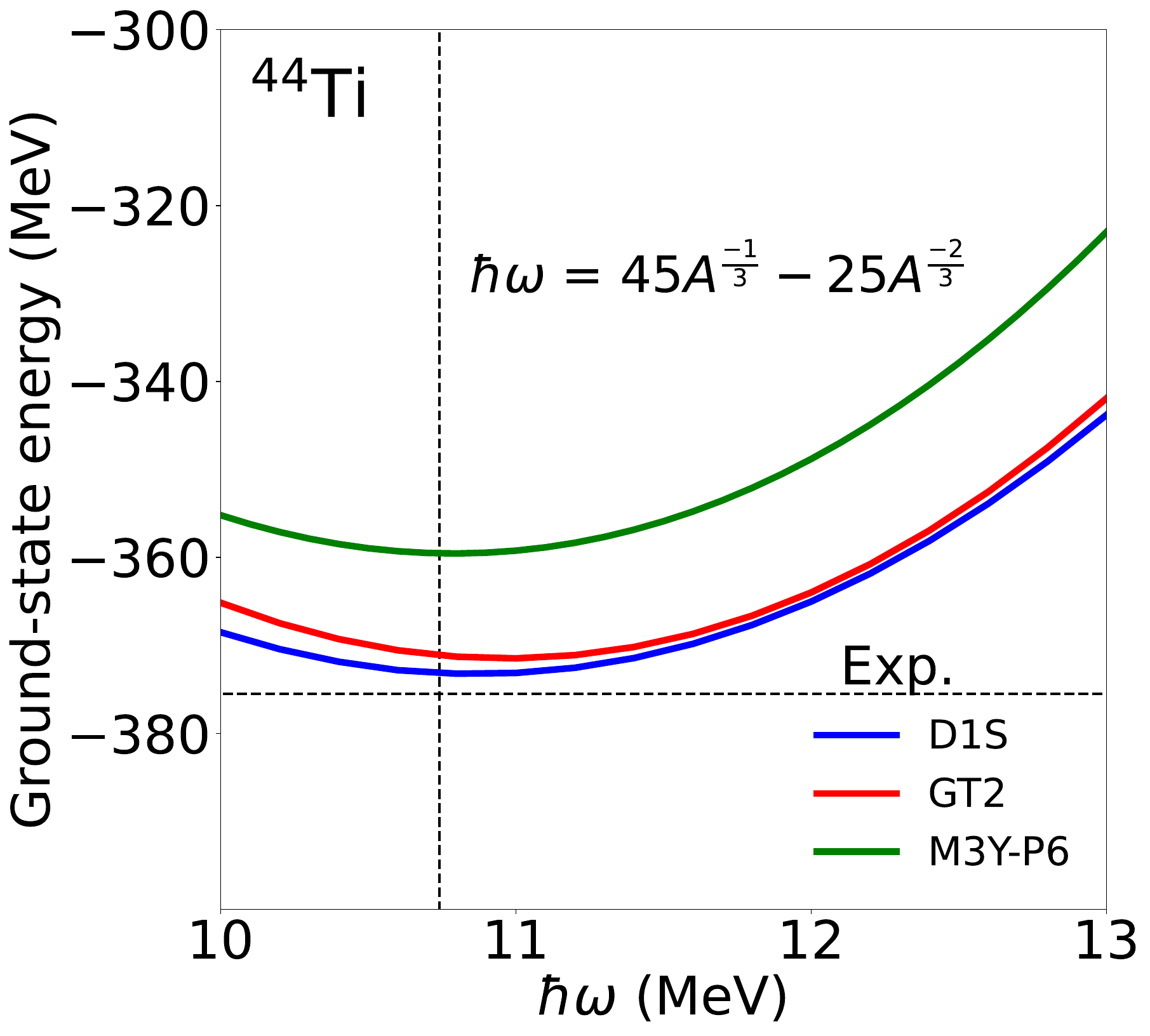}
  \caption{
  Ground-state energy of $^{44}$Ti as a function of the HO energy, $\hbar\omega$. The blue, red, and green solid lines are the SM results obtained with the Gogny-D1S, GT2, and M3Y-P6 functionals. The horizontal and vertical dotted lines denote the experimental value and the empirical HO energy, respectively. }
  \label{fig:gseTi44}
\end{figure}

Before showing numerical results, let us first show the validity of the empirical formula for the HO frequency $\omega$.
Figure~\ref{fig:gseTi44} shows the ground-state energies, Eq.~(\ref{eq:Egs}),
of $^{44}$Ti obtained with the three density-dependent interactions compared with experimental data \cite{Wang2021}.
The HO energy $\hbar \omega = 45 A^{-1/3} - 25 A^{-2/3}$ given by the empirical formula \cite{BLOMQVIST1968545} is reasonably close to the optimal value for all the interactions.
Hereafter, we adopt this formula to determine $\hbar\omega$ throughout this paper.

\subsection{Two-body matrix elements}
\label{sec:TBME}

\begin{figure*}[tbp]
  \centering
  \begin{minipage}[b]{0.3\linewidth}
    \includegraphics[width=\linewidth]{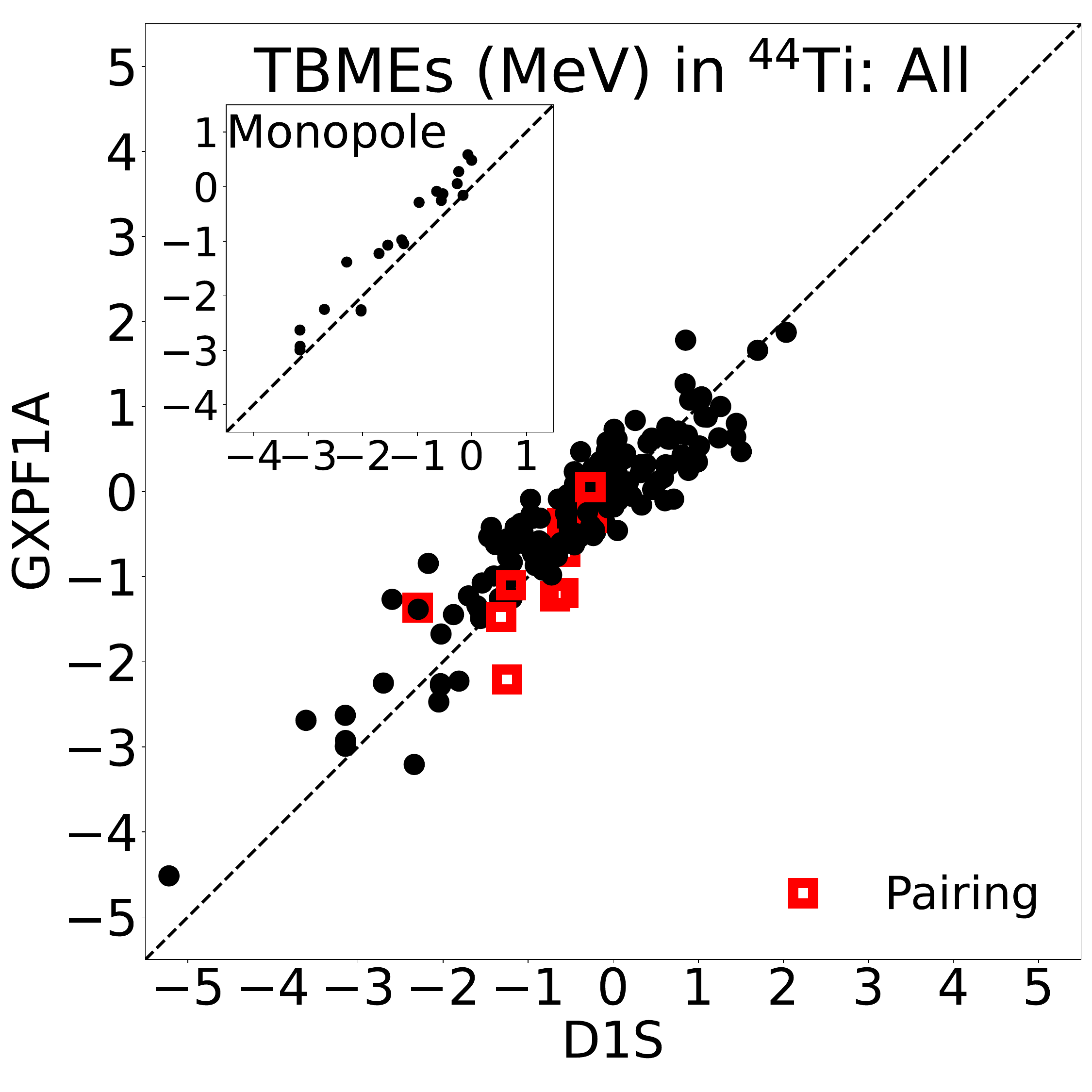}
    \centerline{(a)}
  \end{minipage}
  \hfill
  \begin{minipage}[b]{0.3\linewidth}
    \includegraphics[width=\linewidth]{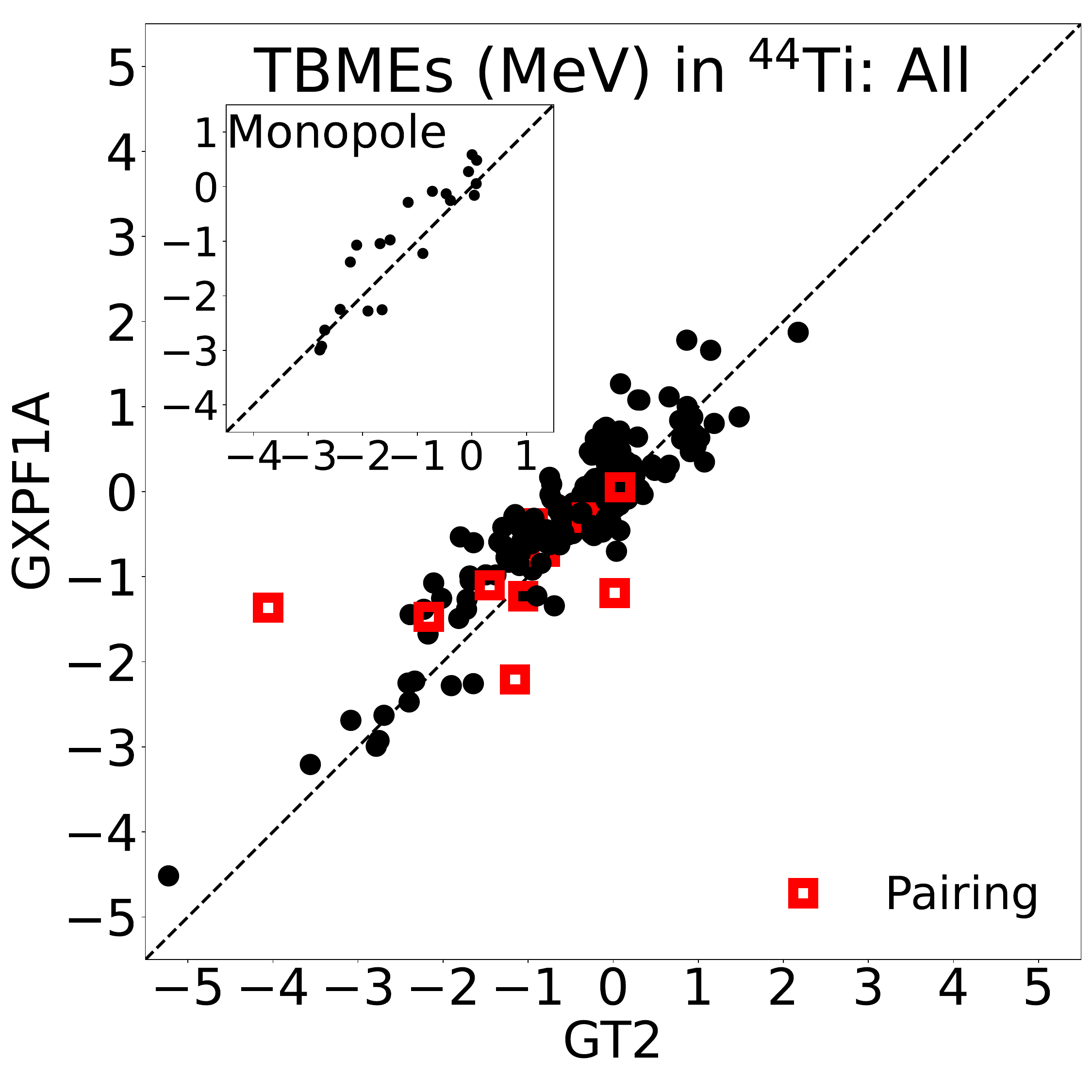}
    \centerline{(b)}
  \end{minipage}
  \hfill
  \begin{minipage}[b]{0.3\linewidth}
    \includegraphics[width=\linewidth]{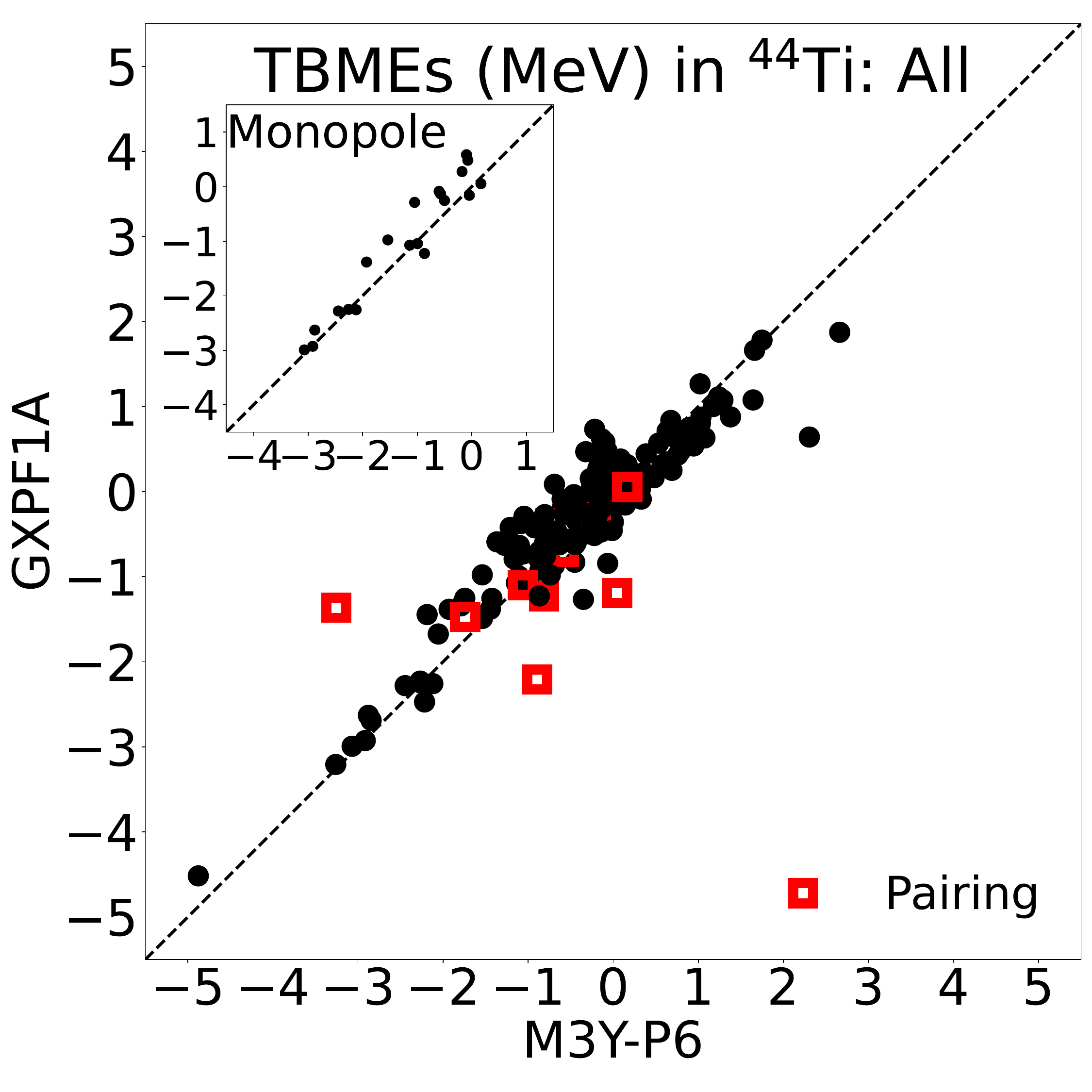}
    \centerline{(c)}
  \end{minipage}

  \caption{Correlation of the all TBMEs of GXPF1A against (a) Gogny-D1S, (b) Gogny-GT2, and (c) M3Y-P6 in $^{44}$Ti. The red open squares denote the $J=0$ matrix elements. The insets show the monopole matrix elements.}
  \label{fig:TBME44Ti}
\end{figure*}

The SM TBMEs are obtained from the Gogny-type and M3Y-type density-dependent interactions of Eqs.~(\ref{eq:gognyint}) and (\ref{eq:m3yint}).
Figure \ref{fig:TBME44Ti} shows the correlation of TBMEs $\langle a,b|V|c,d\rangle_{J,T}$ in $^{44}$Ti between the GXPF1A interaction, as an empirical interaction, and the adopted density-dependent interactions of Gogny-D1S, Gogny-GT2, and M3Y-P6.
The GXPF1A interaction is one of the well-established effective interactions for $pf$-shell nuclei and is based on the chi-square-fitted GXPF1 interaction \cite{PhysRevC.65.061301} with minor revision to reproduce the $N = 32$ shell gap.

Figure \ref{fig:TBME44Ti} shows that all the points are located approximately near the diagonal line,
indicating that the two TBMEs of Gogny-D1S, Gogny-GT2, and M3Y-P6 reasonably agree with those of GXPF1A.
The TBMEs of the GXPF1A interaction have the mass dependence with the factor $(A/42)^{-0.3}$, which we found to be essential for the good agreement of the TBMEs in a broad region of nuclei. 
The differences are within 2 MeV for the Gogny-D1S and M3Y-P6 effective interactions.
The largest deviation is found for
the matrix element $\langle 0f_{7/2},0f_{7/2}|V|1p_{3/2},1p_{3/2}\rangle_{J=0,T=1}$ of Gogny-GT2, -4.056 MeV.
We find a relatively large deviation for the same TBME in M3Y-P6 as well, -3.255 MeV.
This element represents an off-diagonal component of the pairing interaction, and its deviation is clearly larger than that of other elements, including diagonal ones.

The inset figures in Fig. \ref{fig:TBME44Ti} (a-c) show the monopole matrix elements derived from Gogny-D1S, GT2, and M3Y-P6 compared with GXPF1A.
The monopole matrix element $V^{\rm{Monopole}}_{a,b,T}$ is defined
in terms of the TBMEs $\langle a,b|V|c,d\rangle_{J, T}$ as
\begin{align}
    V^{\rm{Monopole}}_{a,b,T} =  \frac{\sum\limits_{\scriptscriptstyle J} (2J+1) \langle a,b|V|a,b\rangle_{J, T}}{\sum\limits_{\scriptscriptstyle J} (2J+1)},
\end{align}
where $J$ is the angular momentum of two-nucleon state ($a,b$) \cite{otsuka2020evolution},
then, averaged over all possible values of $J$.
The monopole interaction plays a crucial role in the variation of the shell structure.
The monopole matrix elements of the GXPF1A interaction in the inset figures of Fig. \ref{fig:TBME44Ti} are very similar to those of Gogny-D1S, GT2, and M3Y-P6, indicating that these theoretical models do not introduce any anomalous contributions to the single-particle energies.

\begin{figure}[htbp]
  \centering
  \includegraphics[width=1.0\linewidth]{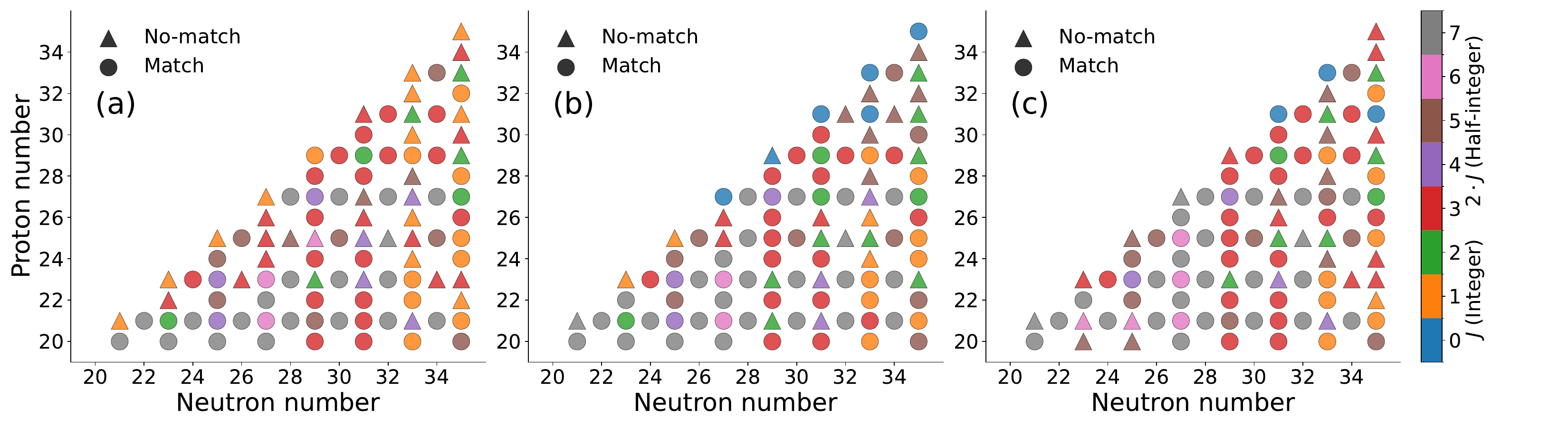}
  \caption{
  Ground-state spin, $J$, given by the SM results with (a) Gogny-D1S, (b) Gogny-GT2, and (c) M3Y-P6 functionals. The spin $J$ (even-mass nuclei) or $2J$ (odd-mass nuclei) is indicated by colors. The circles (triangles) denote the agreement (disagreement) with the experimental data. The even-even nuclei are omitted since their ground-state spin-parities are trivial, $0^+$ without exception.}
  \label{fig:nchart}
\end{figure}

\subsection{Ground state properties \label{sec:gs}}

\begin{figure}[t]
  \centering
  \includegraphics[width=1.0\linewidth]{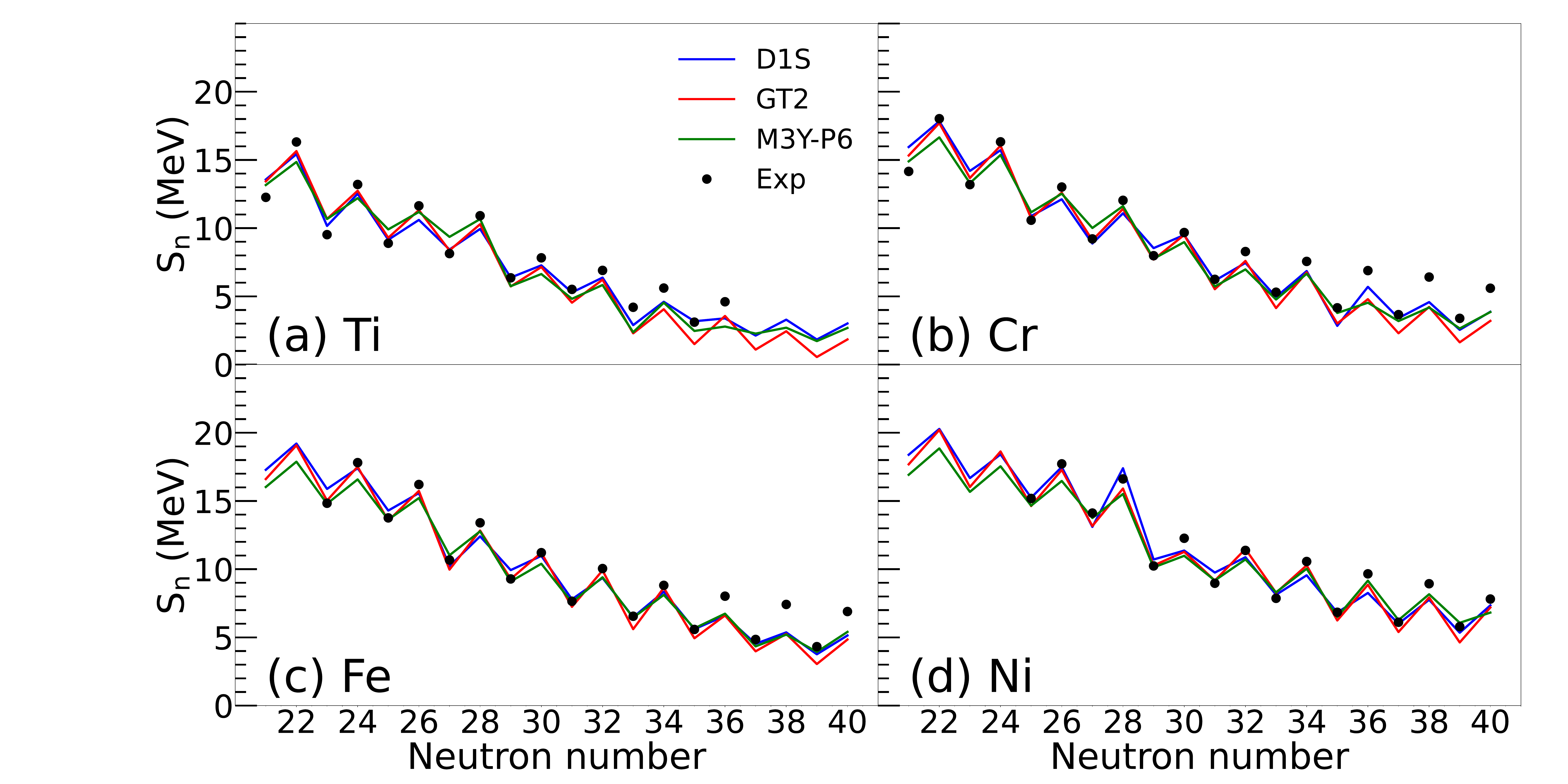}
  \caption{
  One-neutron separation energies of  (a) Ti, (b) Cr, (c) Fe,  and (d) Ni isotopes as a function of the neutron number. The blue, red, and green lines denote the theoretical results with the Gogny-D1S, Gogny-GT2, and M3Y-P6 interactions, respectively. The black circles denote the experimental values taken from \cite{Wang2021}.
  }
  \label{fig:neusepene}
\end{figure}

In Fig.~\ref{fig:gseTi44}, the ground-state energy of $^{44}$Ti is shown.
The Gogny-D1S and GT2 results are similar and in good agreement with the experimental data.
The M3Y-P6 underestimates the binding energy by about 20 MeV.
A possible reason is that the the Yukawa-type functions are less well represented by the HO basis than the Gaussian functions.
In addition, the parameters of M3Y-P6 were determined allowing slight underbinding (Table $\rm{I\hspace{-1.2pt}V}$ in Ref.~\cite{PhysRevC.87.014336}), although the Gogny-D1S overestimates the binding energies (shown in the same table). 
Despite of the underbinding of the ground-state energy, the separation energies are well reproduced, as will be shown in Fig.~\ref{fig:neusepene}.

Figure~\ref{fig:nchart} shows (by colors) the ground-state spin of $pf$-shell even-odd and odd-odd nuclei predicted
by three density-dependent interactions.
One can see (by shapes) whether these predictions are correct or not compared with the experimental data \cite{NNDC_NuDat3}.
In the present study, even-mass nuclei have positive parity and odd-mass nuclei have negative parity, consistent with experimental data.
Each figure shows that the theoretical models reproduce the ground-state spin and parity values reasonably well.
Among the 100 $pf$-shell nuclei we surveyed, the agreement ratios are 63.0\%, 72.0\%, and 69.0\% for the Gogny-D1S, GT2, and M3Y-P6, respectively.
In the $pf$-shell odd-mass nuclei, the agreement ratios of the Gogny-D1S, GT2, and M3Y-P6 interactions are 75.0\%, 81.2\%, and 79.7\%, while the prediction accuracies of the odd-odd nuclei are 41.7\%, 55.6\%, and 50.0\%, respectively. Thus, the ratio of odd-odd nuclei is slightly worse than the odd-mass case.
The predictive ability is apparently low for $N=Z$ odd-odd nuclei because of the competition of $T=0$ and $T=1$ states and the various coupling of unpaired nucleons.
For comparison, we also performed the same benchmark tests for 61 $sd$-shell nuclei and found similar accuracy. The agreement ratios are  78.7\%, 72.1\%, and 78.7\% for  Gogny-D1S, GT2, and M3Y-P6 functionals, respectively.
The accuracy in the corresponding $sd$-shell nuclei with the USDB effective interaction \cite{PhysRevC.74.034315}, which is a widely used interaction for the $sd$-shell region, is 88.5\%, and that of the GXPF1A interaction is 72.0\% for the $pf$-shell nuclei.
The value for the USDB interaction is clearly higher than for the other functionals, whereas that for GXPF1A is roughly the same as the Gogny-GT2 and M3Y-P6 functionals.
We demonstrate that the density-dependent interactions, determined phenomenologically without specific fitting, have good predictions compared with the optimized SM interactions for $sd$- and $pf$-shell nuclei.

Figure~\ref{fig:neusepene} shows the calculated neutron separation energies of the Ti, Cr, Fe, and Ni isotopes compared with the experimental data \cite{Wang2021}.
The odd-even staggering is clearly seen, and the theoretical results are in good agreement with the corresponding experimental data.
The SM results obtained with these functionals are close to each other in the $N\leq 35 $ region. They well reproduce the characteristic energy difference between $^{50}$Ti and $^{51}$Ti, and that between $^{56}$Ni and $^{57}$Ni, reflecting the $N=28$ magicity.
For neutron numbers above $N = 35$, the calculations do not reproduce the experimental data.
This discrepancy is attributed to the restriction of the model space, which does not include cross-shell contributions between nucleons in the $pf$-shell and those in the $sdg$-shell model space.
This discrepancy in the neutron-rich Cr and Fe isotopes is related to the quadrupole collectivity.
The enhancement of quadrupole collectivity in these isotopes near $N = 40$ is attributed to a rapid shape transition from spherical to deformed nuclei with increasing neutron number~\cite{PhysRevC.82.054301}. 
This collectivity is experimentally supported by the excitation energies and $B(E2)$ transition strengths. 
Moreover, SM studies~\cite{PhysRevC.82.054301} have shown that the occupation numbers of the neutron intruder orbits $0g_{9/2}$ and $1d_{5/2}$ increase rapidly in Cr and Fe isotopes as the neutron number changes from $N = 36$ to $N = 42$, based on calculations in the $pf$–$0g_{9/2}$–$1d_{5/2}$ model space. 
These results indicate that particle–hole excitations involving the intruder orbits play a crucial role in the strong quadrupole collectivity in this region.
In addition, previous work~\cite{PhysRevC.93.035805} evaluated the two-neutron separation energies of Cr isotopes and compared experimental data with SM calculations performed in both the restricted $pf$-shell model space and the extended $pf$–$0g_{9/2}$–$1d_{5/2}$ model space. 
The SM results obtained within the $pf$ shell alone significantly underestimate the experimental separation energies around $N = 38$, whereas calculations including the $0g_{9/2}$ and $1d_{5/2}$ orbits show good agreement with the experimental data. 
This clearly demonstrates that quadrupole correlations associated with excitations to the $0g_{9/2}$ and $1d_{5/2}$ orbits make an essential contribution to the binding energies in this region.

The ground-state quantities have also been systematically studied using mean-field methods and can be compared with our approach~\cite{PhysRevC.91.044315,PhysRevC.76.024320}.
Our published work~\cite{particles8020061} reported systematic trends of the ground-state energies for selected $sd$-shell nuclei and Ca isotopes using both the SM and mean-field approaches with the Gogny-D1S interaction. 
The results for the $sd$-shell nuclei and the Ca isotopes are shown in Figs.~2 and~3 of Ref.~\cite{particles8020061}, respectively. 
The SM calculations successfully reproduce the ground-state energies of the $sd$-shell nuclei in the range $N = 8$--16 and of the $^{40-48}$Ca isotopes. 
Moreover, the SM approach provides a better description than the mean-field approach for
the systematic ground-state energies of the $sd$-shell nuclei with increasing neutron number,
although the binding energies tend to be underestimated compared with the experimental data.
The ground-state spins and parities of odd-mass nuclei predicted by the Hartree–Fock–Bogoliubov calculations
with the Gogny-D1S interaction are available in the AMEDEE database~\cite{Hilaire2007}.
We compare the calculated data with those of the experimental data~\cite{NNDC_NuDat3}.
The agreement ratio for the 40 $sd$-shell odd-mass nuclei obtained with this mean-field approach using the Gogny-D1S interaction is 65.0\%, while that for the 64 $pf$-shell odd-mass nuclei is 56.2\%.
In contrast, the corresponding agreement ratios obtained in the present SM approach are significantly higher: 87.5\% for the $sd$-shell nuclei and 75.0\% for the $pf$-shell nuclei.
These results demonstrate that the present approach reproduces the ground-state spins and parities of
the odd-mass nuclei more accurately than the mean-field method.

\begin{figure}[H]
  \centering

  \begin{minipage}[t]{0.6\linewidth}
    \centering
    \includegraphics[width=\linewidth]{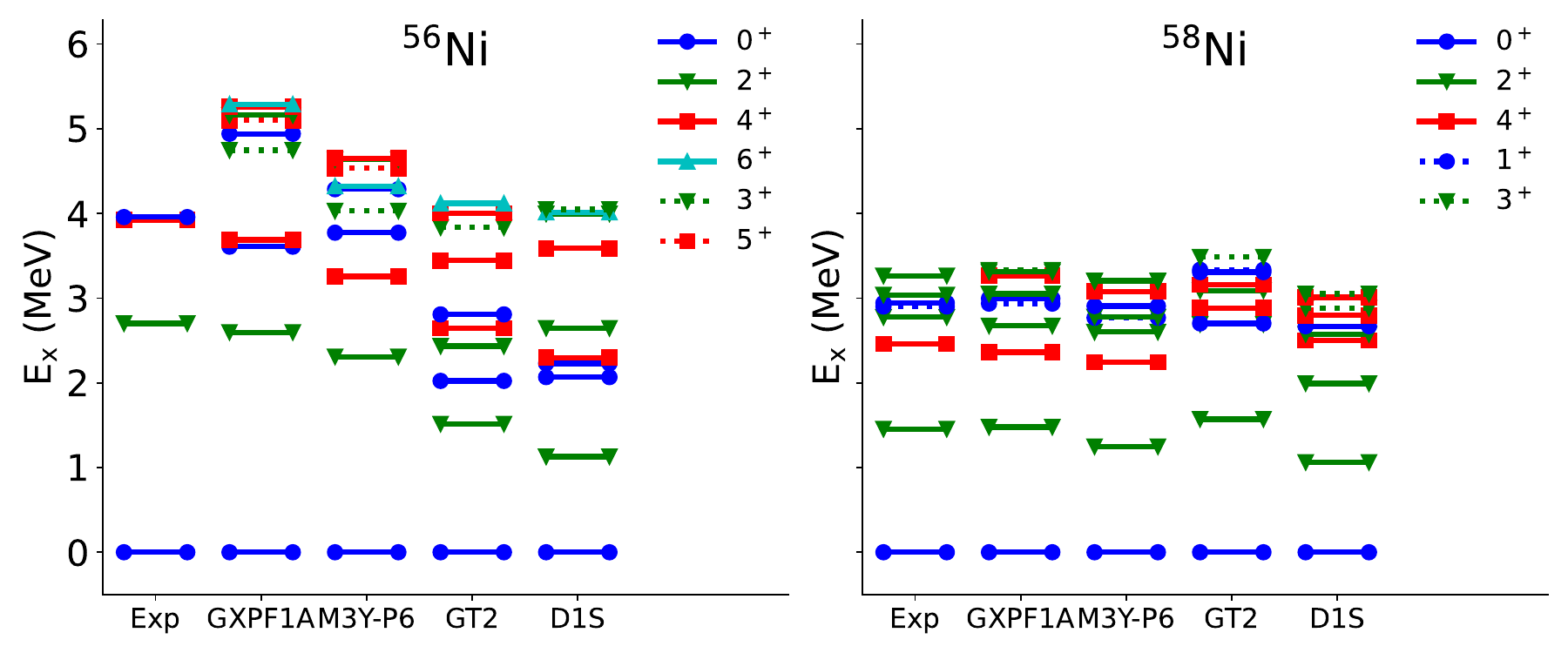}
    \label{fig:nienesubfig1a}
  \end{minipage}

  \begin{minipage}[t]{0.6\linewidth}
    \centering
    \includegraphics[width=\linewidth]{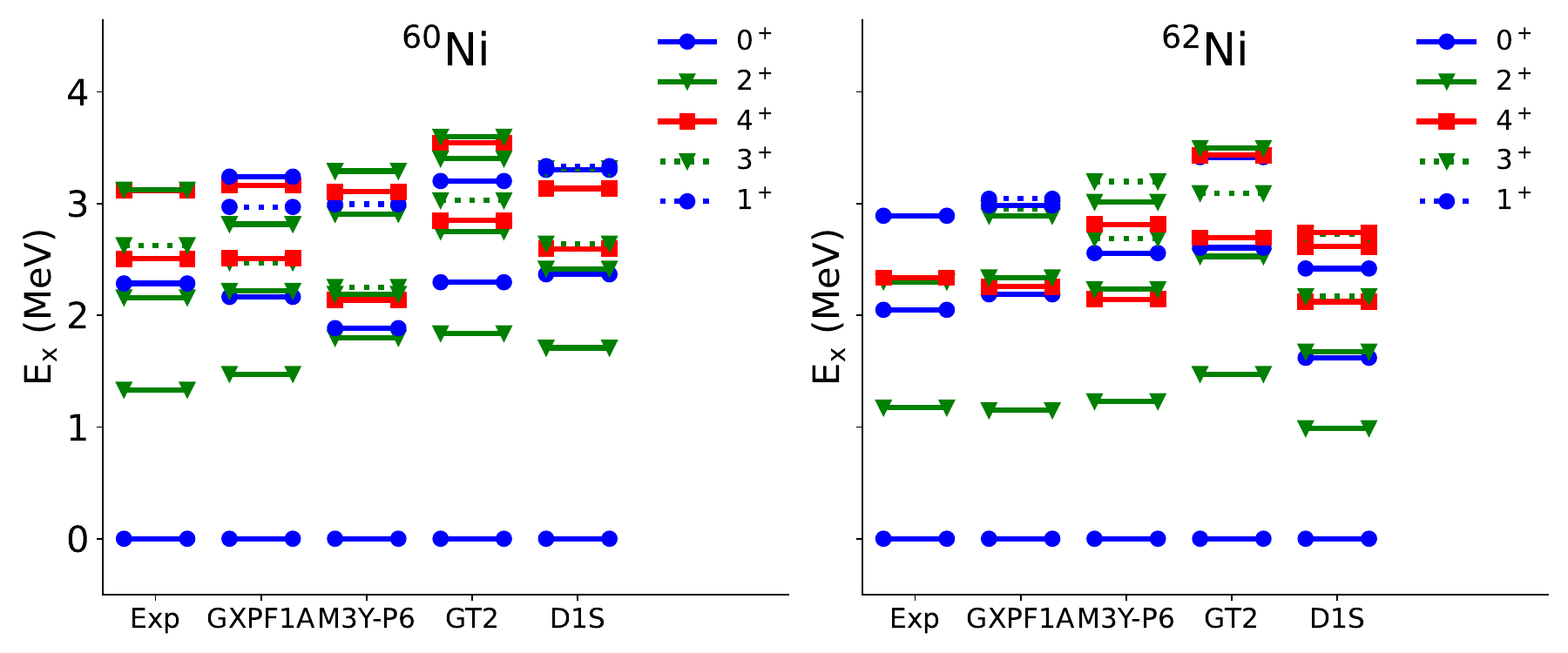}
    \label{fig:nienesubfig1b}
  \end{minipage}

  \begin{minipage}[t]{0.6\linewidth}
    \centering
    \includegraphics[width=\linewidth]{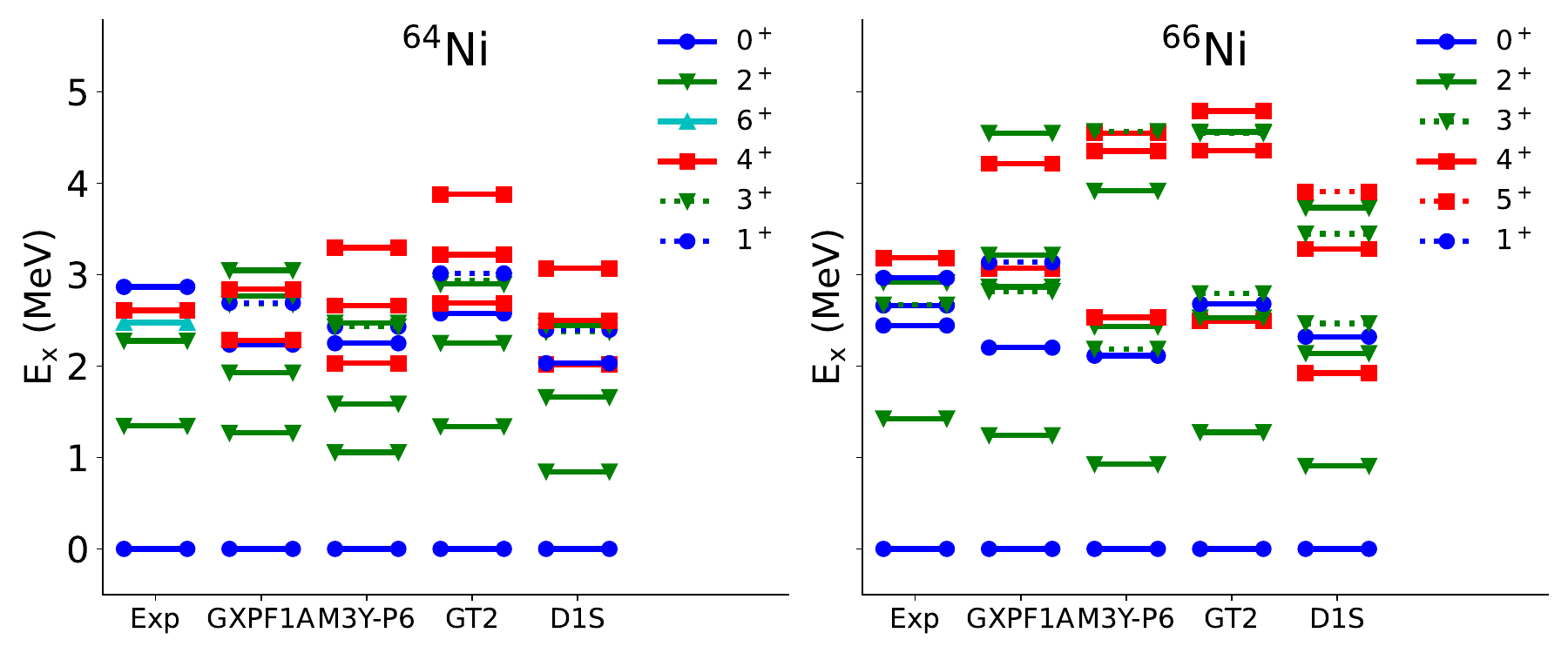}
    \label{fig:nienesubfig1c}
  \end{minipage}

  \caption{Level schemes of even-mass Ni isotopes.
    The experimental values are compared with the SM results obtained using GXPF1A, M3Y-P6, Gogny-GT2, and Gogny-D1S interactions. }
  \label{fig:evenNienespec}
\end{figure}

\subsection{Excited states \label{sec:excited}}

In Sect.~\ref{sec:gs}, the calculations with the density-dependent interactions provide a good description of the binding energies, spin, and parity for the $pf$-shell nuclei.
It is also important to calculate not only ground states but also excited states to further analyze the underlying shell structure and the performance of the models.
Low-lying excitation spectra are important observables, and their changes with neutron number provide insights into the evolution of the nuclear structure.
Figure \ref{fig:evenNienespec} shows the theoretical energy spectra in the even-mass $^{56-66}$Ni isotopes compared with the experimental data.
The Gogny-D1S and GT2 results for the excited states of the $^{58}$Ni and $^{62}$Ni isotopes are in good agreement with the experimental data; however, the $2^+_1$ excitation energy of  $^{56}$Ni is underestimated by these two functionals.
Experimental results for $^{56}$Ni exhibit a relatively high excitation energy of the first excited state,
indicating the magicity of $N=Z=28$.

\begin{figure}[t]
  \centering

  \begin{minipage}[t]{0.9\linewidth}
    \centering
    \includegraphics[width=\linewidth]{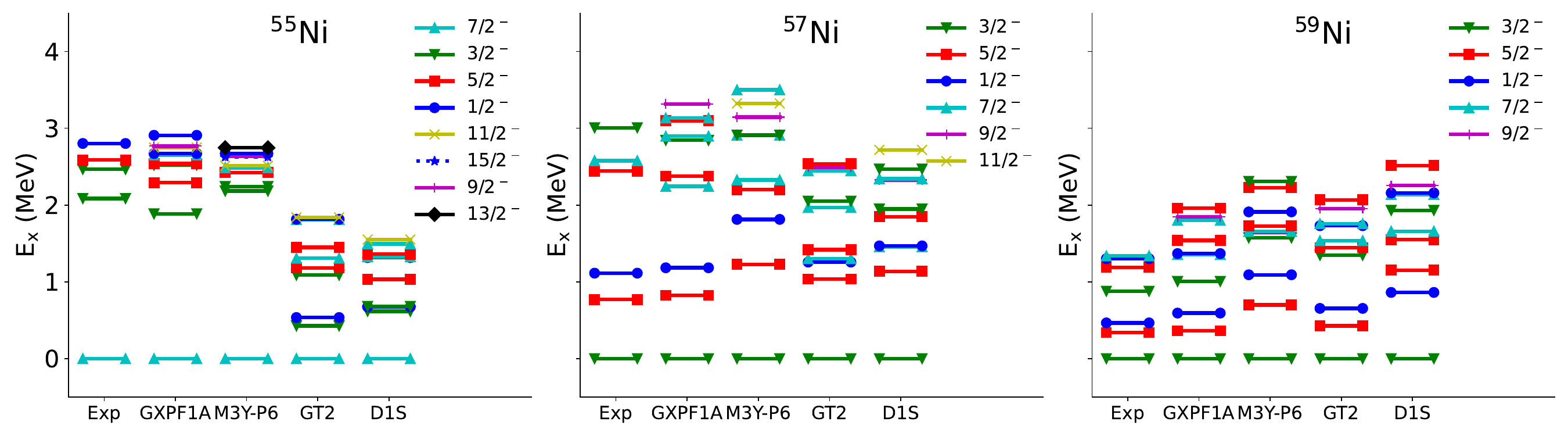}
    \label{fig:nienesubfig1a}
  \end{minipage}

  \begin{minipage}[t]{0.6\linewidth}
    \centering
    \includegraphics[width=\linewidth]{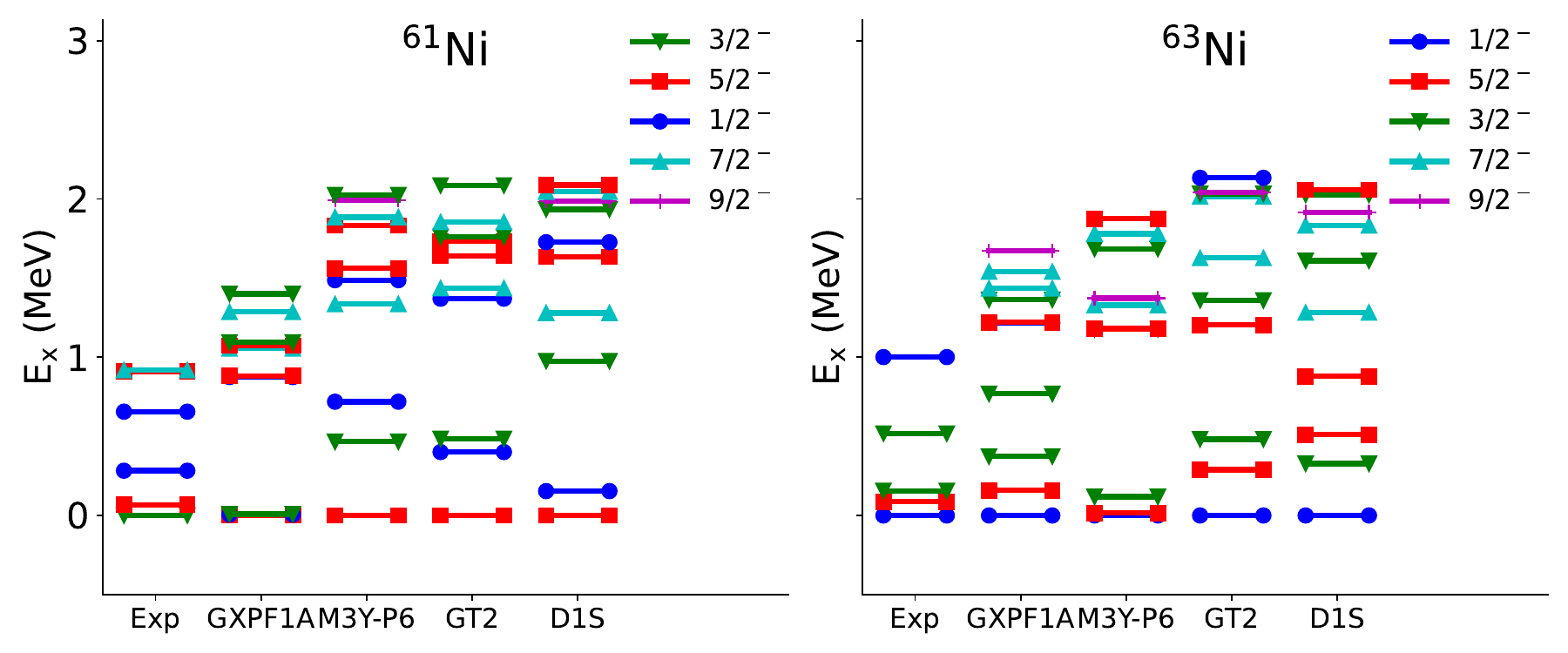}
    \label{fig:nienesubfig1b}
  \end{minipage}
  
  \begin{minipage}[t]{0.6\linewidth}
    \centering
    \includegraphics[width=\linewidth]{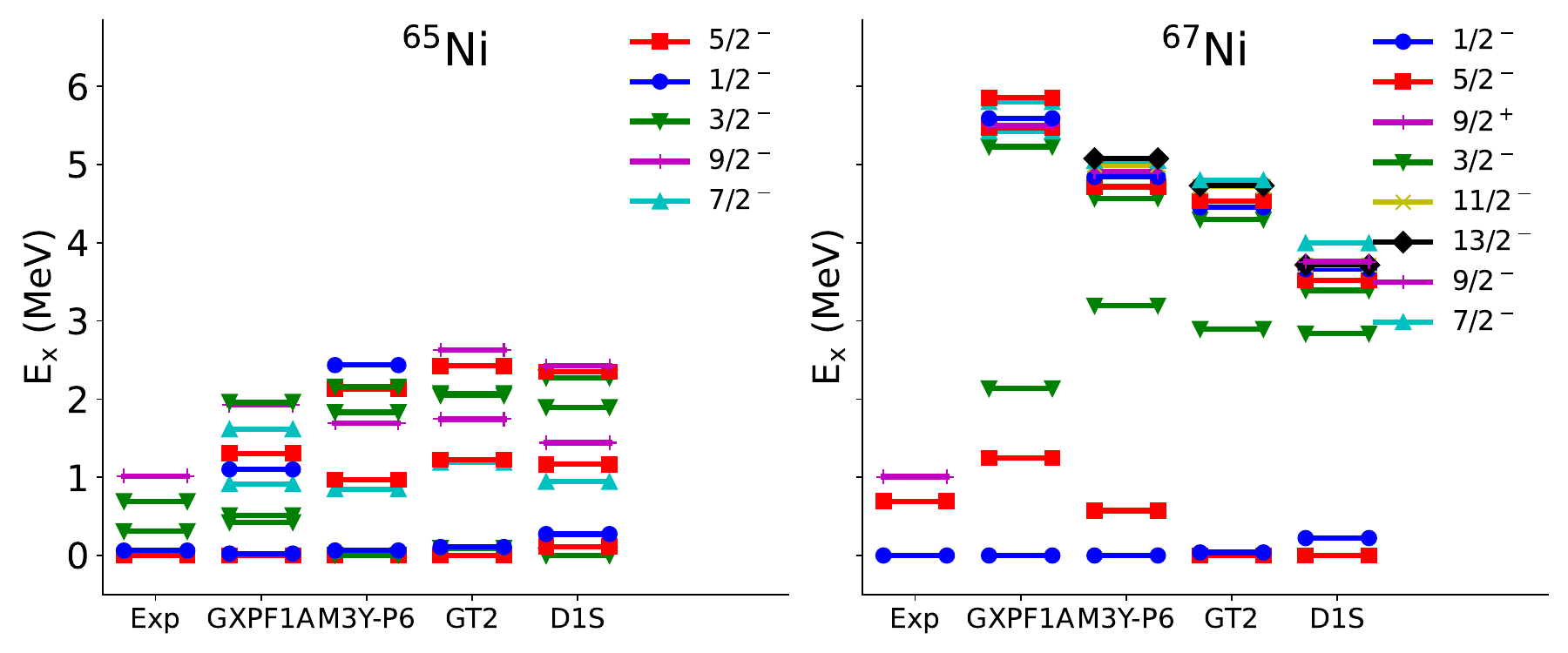}
    \label{fig:nienesubfig1c}
  \end{minipage}

  \caption{Level schemes of odd-mass Ni isotopes. Details are seen in the caption of Fig.~\ref{fig:evenNienespec}.}
  \label{fig:oddNienespec}
\end{figure}

Figure~\ref{fig:oddNienespec} shows the energy spectra of odd-mass Ni isotopes.
For $^{55}$Ni, the energy levels obtained from the experimental data, GXPF1A, and M3Y-P6 interactions show a clearly different behavior from those calculated using the Gogny-D1S and GT2 interactions. 
The excitation energies given by Gogny-D1S and GT2 are significantly smaller than the experimental values and those obtained with GXPF1A and M3Y-P6. 
The latter interactions describe the $N = 28$ magicity, whereas the former fail to do.
The first excitation energies of $^{57}$Ni calculated with Gogny-D1S and GT2 are comparable to the experimental data and the GXPF1A results, although the $2_1^+$ excitation energy is significantly underestimated for $^{56}$Ni.
In $^{61,63}$Ni, inversions of ground-state spins are observed with only a small energy difference.

Among the three functionals, the M3Y-P6 results reproduce the experimental data relatively well, with an accuracy comparable to that of GXPF1A.
The M3Y-P6 interaction provides a good description of the low-lying excitation energies for even-mass and odd-mass Ni isotopes, whereas Gogny-D1S and GT2 fail for $^{55}$Ni and $^{56}$Ni.
This indicates that the former provides a better description of the shell structure for the Ni isotopes than the latter.

\begin{figure}[t]
  \centering

  \begin{minipage}[t]{0.48\linewidth}
    \centering
    \includegraphics[width=\linewidth]{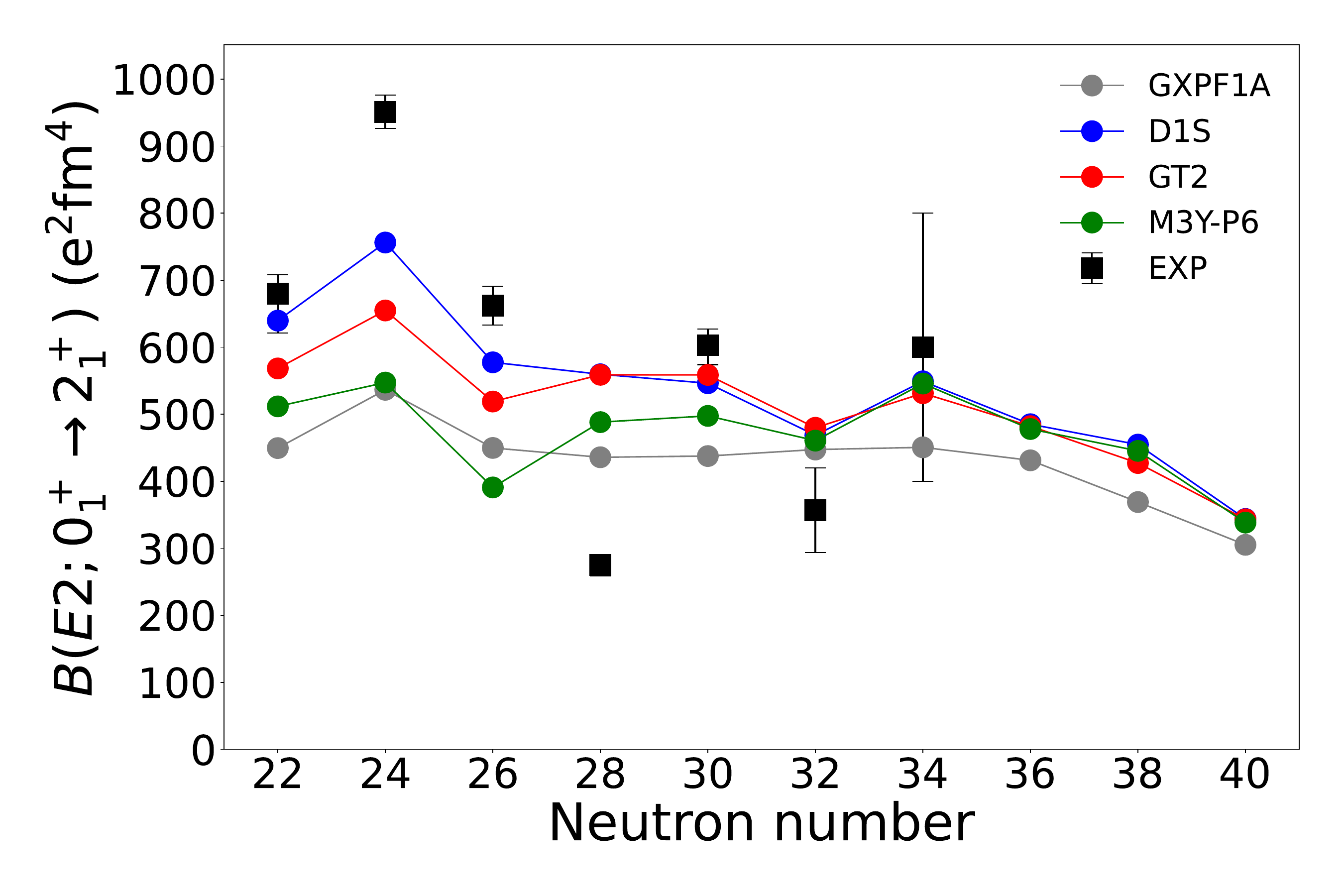}
    \subcaption{}
    \label{fig:subfig1a}
  \end{minipage}
  \hfill
  \begin{minipage}[t]{0.48\linewidth}
    \centering
    \includegraphics[width=\linewidth]{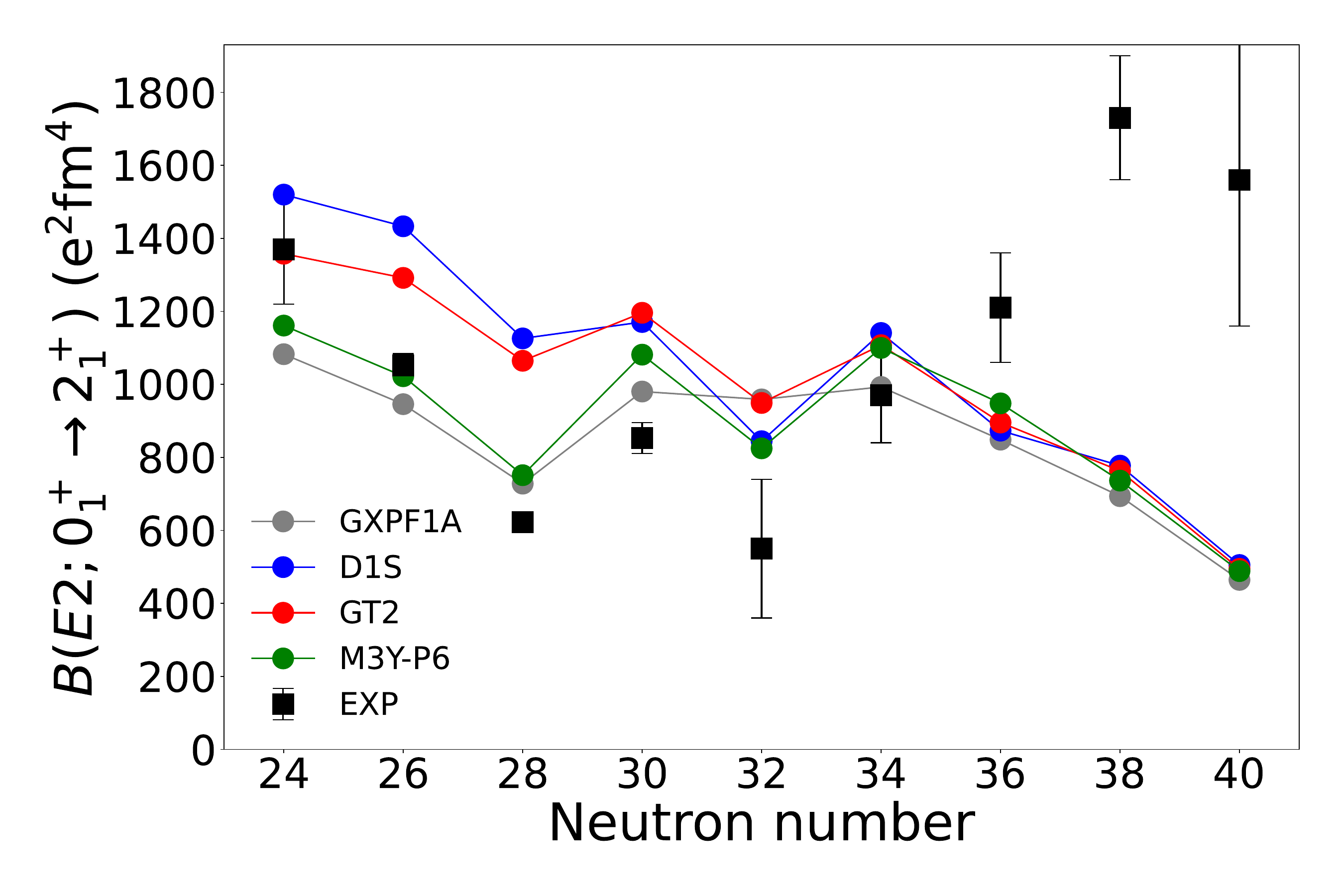}
    \subcaption{}
    \label{fig:subfig1b}
  \end{minipage}

  \vspace{1em}

  \begin{minipage}[t]{0.48\linewidth}
    \centering
    \includegraphics[width=\linewidth]{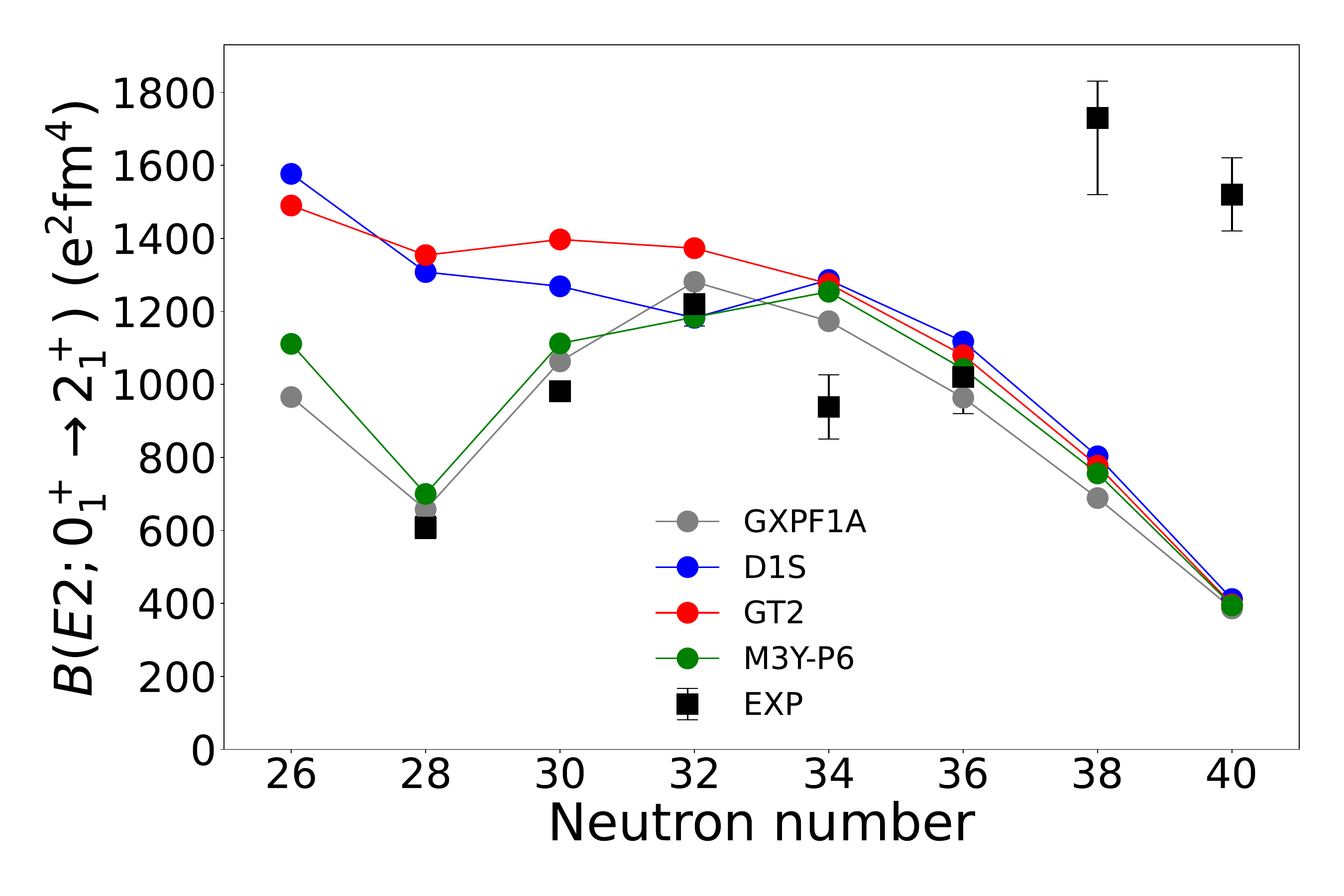}
    \subcaption{}
    \label{fig:subfig1c}
  \end{minipage}
  \hfill
  \begin{minipage}[t]{0.48\linewidth}
    \centering
    \includegraphics[width=\linewidth]{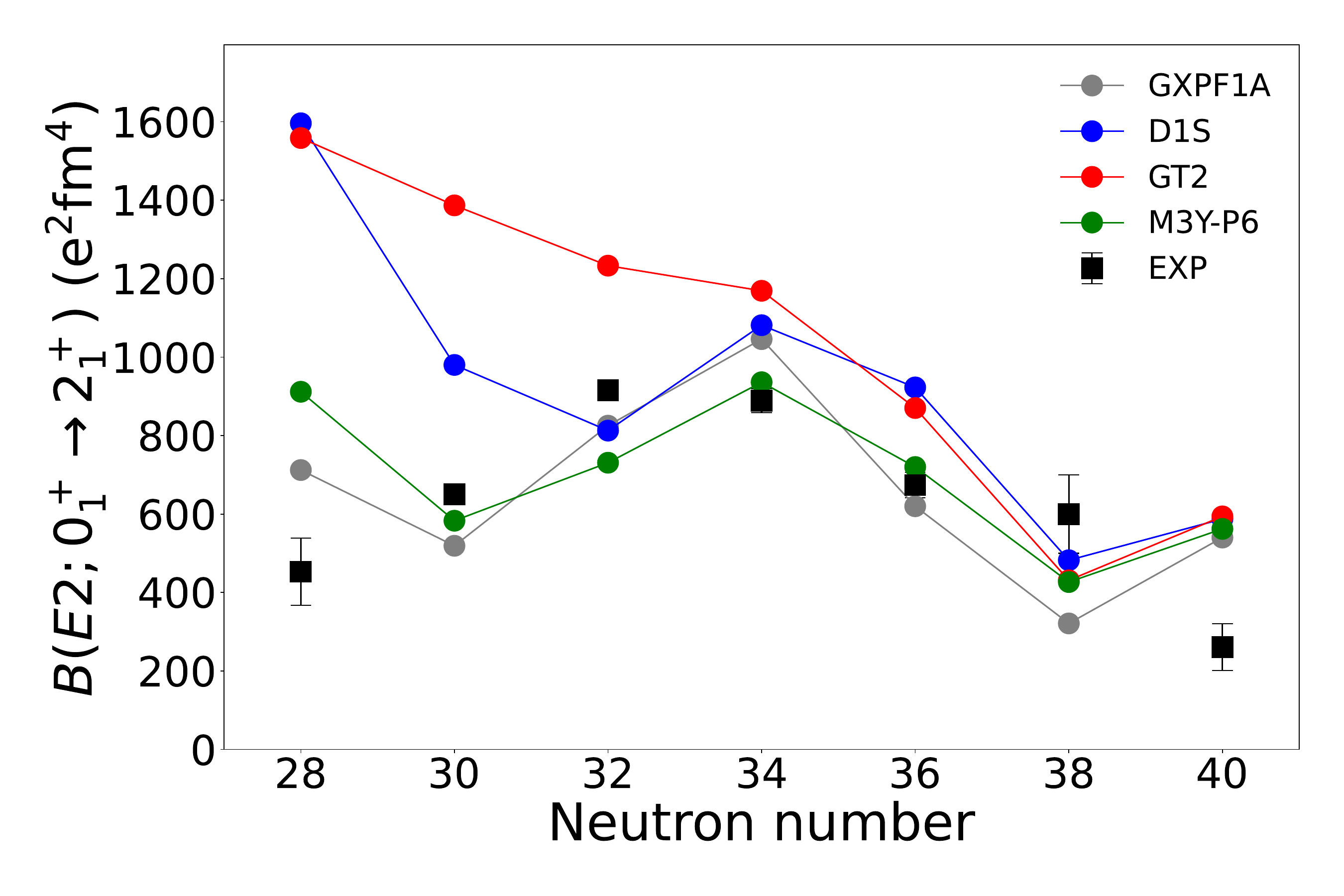}
    \subcaption{}
    \label{fig:subfig1d}
  \end{minipage}

  \caption{Reduced transition probabilities $B(E2;0^+_1 \rightarrow 2^+_1$) of (a) Ti, (b) Cr, (c) Fe, and (d) Ni isotopes. The gray, blue, red, and green symbols with the solid lines denote the SM results with the GXPF1A, Gogny-D1S, Gogny-GT2, and M3Y-P6 interactions, respectively. The HO energy for the GXPF1A interaction is $\hbar\omega = 41A^{-1/3}$ MeV~\cite{Honma2005}.
  The black squares with the error bars show the experimental values taken from Ref.~\cite{PRITYCHENKO20161}.}
  \label{fig:be2}
\end{figure}

The reduced transition probability $B(E2; 0^+_1 \rightarrow 2^+_1)$ is an important measure for discussing quadrupole collectivity.
Figure \ref{fig:be2} shows the $B(E2)$ values of Ti, Cr, Fe, and Ni isotopes calculated with the standard effective charges $(e_p, e_n) = (1.5, 0.5)e$ compared with experimental data \cite{PRITYCHENKO20161}.
The calculated $B(E2)$ strengths agree reasonably well with the GXPF1A result.
The underestimation seen at small neutron numbers, $N=22,24,26$, for Ti isotopes
may be caused by neglecting core-excitation effects.
For the Cr, Fe, and Ni isotopes, the $E2$ strengths calculated with these three functionals reasonably reproduce the experimental data, except for the strong enhancements on $^{60,62,64}$Cr and $^{64,66}$Fe, near the end of the $pf$ shell.
This is because the quadrupole deformation accompanying the excitation to the neutron $0g_{9/2}$ orbit plays a crucial role in these nuclei, and these nuclei are considered to be ``second island of inversion'' \cite{PhysRevLett.120.232501,PhysRevC.82.054301,shimizu2012new}.
Among these functionals, we find that M3Y-P6 is the most successful,
especially for the $N=28$ magicity at $^{54}$Fe and $^{56}$Ni.
For instance, the small $B(E2)$ value for $^{54}$Fe ($N=28$) is well reproduced.
Although a reduction in $B(E2)$ for $^{56}$Ni is not perfectly reproduced,
it is comparable to the empirical interaction GXPF1A and significantly better than the Gogny-D1S and GT2.
The significant overestimation of the $B(E2)$ values with two Gogny interactions may indicate incorrect deformation,
which will be discussed later. 
Note that the KB3G interaction~\cite{POVES2001157}, which is another well-established phenomenological interaction, also gives a similar overestimation of the $B(E2)$ at the $^{56}$Ni to the GXPF1A case \cite{Loelius_Ni_BE2_PhysRevC.94.024340}.

\subsection{Magicity of $^{56}$Ni \label{sec:magicity}}

The M3Y-P6 interaction describes the low-lying spectra of Ni isotopes and the $B(E2)$ values of $pf$-shell nuclei better than those of the Gogny-D1S and GT2 interactions.
In particular, its $B(E2)$ strengths for the $N = 28$ isotones Cr, Fe, and Ni are in good agreement with the experimental data with the same accuracy as those of the GXPF1A interaction.
Since the experimental data for $^{56}$Ni strongly suggests its magicity,
we examine an origin of the failure of Gogny-D1S and GT2 for $^{56}$Ni.

The closed-shell character of $^{56}$Ni is examined by the particle-hole excitation across the $N = Z = 28$ energy gap.
The fractions of the closed-shell configuration in the ground-state wave functions
are 27.7\% and 16.6\% for the Gogny-D1S and GT2 interactions, respectively. 
On the other hand, the fractions of the closed configuration for the M3Y-P6 and GXPF1A interactions
are 46.4\% and 67.9\%, which are relatively high and indicate the magicity of $^{56}$Ni,
leading to the relatively large $2^+$ excitation energy shown in Fig.~\ref{fig:evenNienespec}. 

\begin{figure}[htbp]
  \centering
  \includegraphics[width=1.0\linewidth]{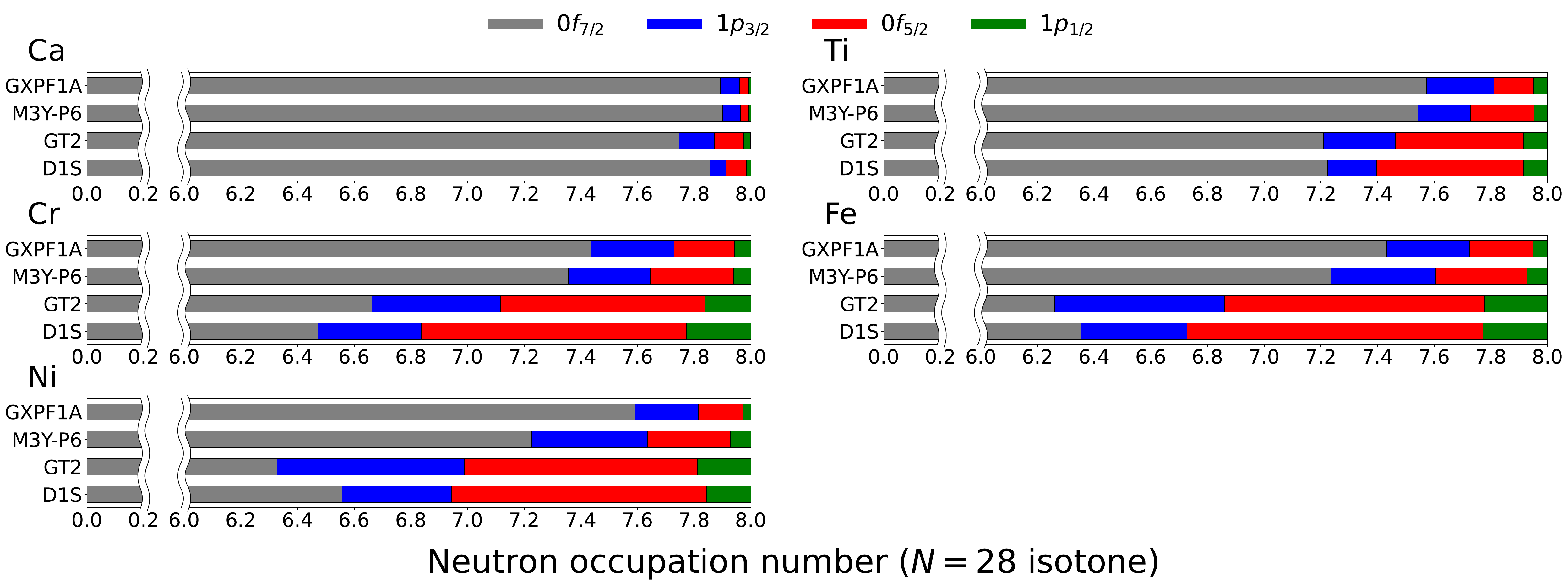}
  \caption{Occupation numbers of the neutron $0f_{7/2}$, $0f_{5/2}$, $1p_{3/2}$, and $1p_{1/2}$ orbits of $N=28$ isotones Ca, Ti, Cr, Fe, and Ni. 
  The gray, red, blue, and green bars correspond to the $0f_{7/2}$, $0f_{5/2}$, $1p_{3/2}$, and $1p_{1/2}$ orbits, respectively.
  The theoretical results are obtained with the GXPF1A, M3Y-P6, GT2, and D1S interactions.
  }
  \label{fig:occnumneu28}
\end{figure}

Figure~\ref{fig:occnumneu28} shows the neutron occupation numbers in $N = 28$ isotones in $pf$-shell orbits.
The neutron occupation numbers in $1p_{3/2}$ and $0f_{5/2}$ obtained with the Gogny-D1S and GT2 interactions are larger than those obtained with the GXPF1A and M3Y-P6 interactions.
A wrong breaking of the magicity at $N = Z = 28$ for Gogny-D1S and Gogny-GT2
is suggested by the excitation spectra in $^{56}$Ni,
the neutron-occupation number shown in Fig. \ref{fig:occnumneu28},
and the calculated $B(E2)$ values.

\begin{figure}[htbp]
  \centering
  \includegraphics[width=1.0\linewidth]{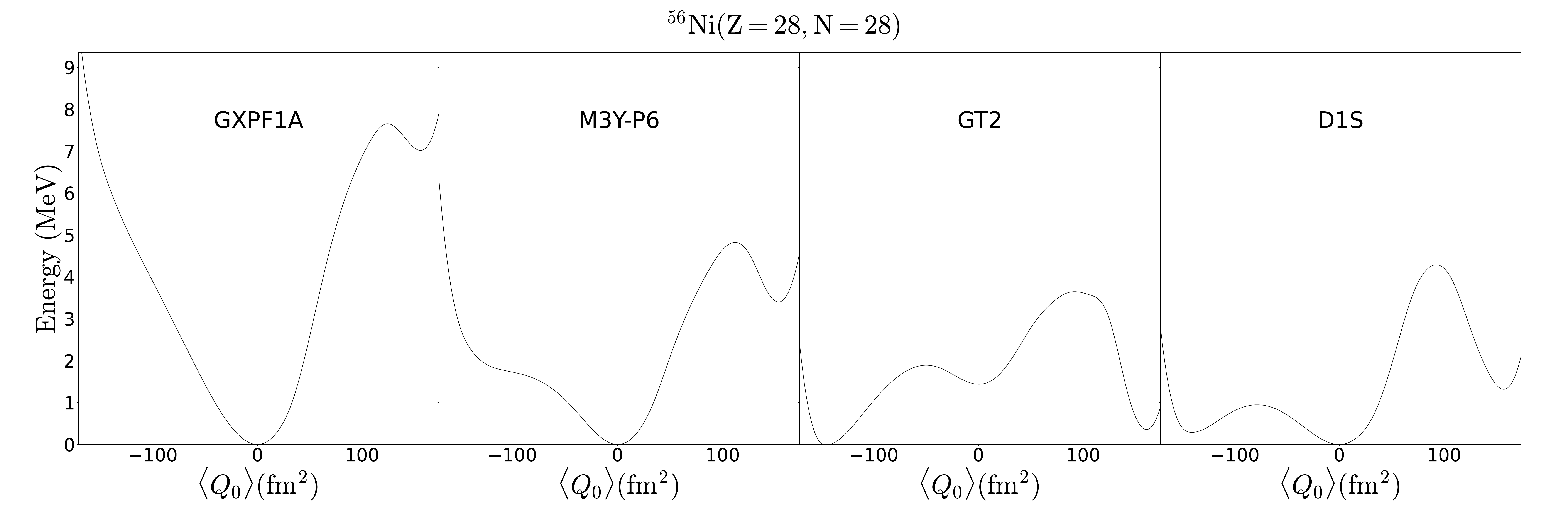}
  \caption{
  Energy surfaces of $^{56}$Ni as a function of the mass quadrupole moment $\langle Q_{0} \rangle $.
  The four panels show the results with the GXPF1A, M3Y-P6, Gogny-GT2, and D1S interactions from left to right.
  }
  \label{fig:pes_ni56}
\end{figure}

Figure~\ref{fig:pes_ni56} shows the energy surfaces of $^{56}$Ni relative to its minimum,
calculated with the Gogny-D1S, GT2, M3Y-P6, and GXPF1A interactions.
These results are obtained with the Hartree-Fock calculations with a constraint of the mass quadrupole moment employing the SM Hamiltonian in the $pf$-shell model space.
The energy surfaces of the GXPF1A and the M3Y-P6 interactions obviously show that the ground state is spherical.
In contrast, the Gogny-D1S interaction exhibits two local minima at energies comparable to the spherical minimum, while the Gogny-GT2 interaction clearly shows that the ground state has an oblate deformation. 
Although the paring effect is neglected in the potential surface,
a qualitative feature on the quadrupole correlation of the adopted interactions
is represented in Fig.~\ref{fig:pes_ni56}.
Reference~\cite{PhysRevC.59.R1846} shows that $^{56}$Ni exhibits a triple shape coexistence
with prolate and oblate bands, in addition to the spherical ground state.
The surface of the M3Y-P6 interaction shows a spherical minimum, a shallow slope region at the oblate side, and a local minimum at the prolate deformation. These features are consistent with the interpretation of the triple shape coexistence.

\section{Summary}
\label{sec:summary}

We performed the SM calculations of $pf$-shell nuclei with the $0\hbar\omega$ model space, employing three types of finite-range density-dependent interactions: Gogny-D1S, Gogny-GT2, and M3Y-P6. 
Without any further phenomenological correction, the SM calculations reproduce the experimental data, including the neutron-separation energies, low-lying spectra, and $E2$ transition probabilities reasonably,
except for the $N>36$ Cr and Fe isotopes, in which the $0g_{9/2}$ orbit plays a crucial role, causing large deformation \cite{PhysRevLett.120.232501,PhysRevC.82.054301,shimizu2012new}. 
The present models reproduce the ground-state spins of the $pf$-shell nuclei with an accuracy comparable to that of the well-established GXPF1A interaction. 
Although these three functionals work reasonably well,
the M3Y-P6 interaction is the best in respect to the correct description of
the magicity of $N=Z=28$ in $^{56}$Ni and its neighboring isotopes. 

We have performed a preliminary study of large-scale SM calculations including the $0g_{9/2}$ orbit to describe the large quadrupole deformation around $^{64}$Cr, but could not reproduce it. 
A previous SM study using the Gogny-D1S interaction also failed to describe the rotational band of $^{30}$Ne in the island of inversion with the $sdpf$-shell model space \cite{Liu_2025}. 
In the current framework, the $LS$-closed shell gap would be too large to describe these islands of inversion. This problem and the fragile $N=Z=28$ gap discussed in Sect.~\ref{sec:magicity} might be remedied simultaneously by introducing a strong isovector spin-orbit force \cite{PhysRevLett.126.172503,yue2024prex}. 
Further investigations along this direction are in progress.

\section*{Acknowledgment}

We acknowledge Yusuke Tsunoda and Hitoshi Nakada for fruitful discussions, and Daisuke Abe for his HFB code.
This work is supported by JST SPRING Grant No. JPMJSP2124 and JST ERATO Grant No. JPMJER2304,
and also by KAKENHI Grant No. JP23K25864, JP24H00239, JP25K00995, and JP25H01268  .
N.S. acknowledges the support of the ``Program for promoting research on
the supercomputer Fugaku'', MEXT, Japan (JPMXP1020230411).
The numerical calculations were performed mainly at the Miyabi-C and Pegasus supercomputers (MCRP program at CCS, University of Tsukuba: NUCLSM, NUCLDFT, and wo22i002).


\bibliographystyle{ptephy}

\bibliography{gognysm_ptep}
%



\appendix

\section{\label{app:paramlist}Parameters of density functionals}

In this appendix, we present the parameter sets of the density functionals we used in the present study: Gogny-D1S \cite{BERGER1991365}, Gogny-GT2 \cite{PhysRevLett.97.162501,dabe_thesis}, and M3Y-P6 \cite{PhysRevC.87.014336}.
These parameters are summarized in Tables \ref{tab:gognyparams} and \ref{tab:m3yparams}.

The Gogny-type interaction \cite{PhysRevC.21.1568} was originally introduced to reproduce the nuclear properties of spherical nuclei using a self-consistent approach.
The Gogny-D1S interaction reproduces the ground-state properties of many nuclei in a broad mass region.
To reproduce the shell evolution of exotic nuclei,
the Gogny-GT2 interaction was proposed to incorporate
the tensor force added to the original Gogny-type interaction
\cite{PhysRevLett.97.162501}.

The M3Y interaction was based on the $G$-matrix elements from the Reid-Elliott soft-core nucleon-nucleon interaction \cite{Bertsch1977M3Y}.
M3Y-P0 denotes the original M3Y-type interaction in Ref.~\cite{PhysRevC.68.014316} without the density-dependent force, fitted for the $G$-matrix interaction derived from the Paris nucleon-nucleon potential \cite{PhysRevC.68.014316}.
One of the latest parameterized interactions is the M3Y-P6 interaction \cite{PhysRevC.87.014336}, whose parameter set was fitted again for the new $G$-matrix interaction by comparing microscopic calculation results for the neutron-matter energy and the binding energy of $^{100}$Sn \cite{FRIEDMAN1981502}.

\begin{table}[t]
  \caption{Parameters of the Gogny-D1S and GT2 interactions.}
  \label{tab:gognyparams}
  \centering
  \begin{tabular}{lcc}
    Parameter set & D1S & GT2 \\
    \hline
    \multicolumn{3}{l}{\textbf{Central force}} \\
    $\mu^{(\mathrm{C})}_1$ (fm) & 0.7 & 0.7 \\
    $t^{(\mathrm{SE})}_1$ (MeV) & 190.83 & 29 \\
    $t^{(\mathrm{TE})}_1$ (MeV) & -836.25 & -1007 \\
    $t^{(\mathrm{SO})}_1$ (MeV) & -6231.45 & 11553 \\
    $t^{(\mathrm{TO})}_1$ (MeV) & -4.37 & -1331 \\
    \hline
    $\mu^{(\mathrm{C})}_2$ (fm) & 1.2 & 1.2 \\
    $t^{(\mathrm{SE})}_2$ (MeV) & -119.60 & -97 \\
    $t^{(\mathrm{TE})}_2$ (MeV) & -120.96 & -61 \\
    $t^{(\mathrm{SO})}_2$ (MeV) & 653.84 & -1357 \\
    $t^{(\mathrm{TO})}_2$ (MeV) & 1.28 & 159 \\
    \hline
    \multicolumn{3}{l}{\textbf{Spin-orbit force}} \\
    $W_{\rm{LS}}$ (MeV\,fm$^{5}$) & 130 & 160 \\
    \hline
    \multicolumn{3}{l}{\textbf{Tensor force}} \\
    $\mu_{\mathrm{TS}}$ (fm) & --- & 1.2 \\
    $V_{\rm{TS}}$ (MeV) & --- & 50.8 \\

    \hline
    \multicolumn{3}{l}{\textbf{Density-dependent force}} \\
    $t_3$ (MeV\,fm$^{3+3\alpha}$) & 1390.6 & 1400.0 \\
    $x_3$ & 1.0 & 1.0 \\
    $\alpha$ & $1/3$ & $1/3$ \\
  \end{tabular}

\end{table}

\begin{table}[t]
  \caption{Parameter set of M3Y-type interactions.}
  \label{tab:m3yparams}
  \centering
  \scalebox{0.9}{
  \begin{tabular}{lcc}
    Parameter set & M3Y-P0 & M3Y-P6 \\
    \hline
    \multicolumn{3}{l}{\textbf{Central force}} \\
    $\mu^{(\mathrm{C})}_1$ (fm) & 0.25 & 0.25 \\
    $t^{(\mathrm{SE})}_1$ (MeV) & 11466 & 10766 \\
    $t^{(\mathrm{TE})}_1$ (MeV) & 13967 & 8474 \\
    $t^{(\mathrm{SO})}_1$ (MeV) & $-1418$ & $-728$ \\
    $t^{(\mathrm{TO})}_1$ (MeV) & 11345 & 12453 \\
    $\mu^{(\mathrm{C})}_2$ (fm) & 0.40 & 0.40 \\
    $t^{(\mathrm{SE})}_2$ (MeV) & $-3556$ & $-3520$ \\
    $t^{(\mathrm{TE})}_2$ (MeV) & $-4594$ & $-4594$ \\
    $t^{(\mathrm{SO})}_2$ (MeV) & 950 & 1386 \\
    $t^{(\mathrm{TO})}_2$ (MeV) & $-1900$ & $-1588$ \\
    $\mu^{(\mathrm{C})}_3$ (fm) & 1.414 & 1.414 \\
    $t^{(\mathrm{SE})}_3$ (MeV) & $-10.463$ & $-10.463$ \\
    $t^{(\mathrm{TE})}_3$ (MeV) & $-10.463$ & $-10.463$ \\
    $t^{(\mathrm{SO})}_3$ (MeV) & 31.389 & 31.389 \\
    $t^{(\mathrm{TO})}_3$ (MeV) & 3.488 & 3.488 \\
    \hline
    \multicolumn{3}{l}{\textbf{Spin-orbit force}} \\
    $\mu^{(\mathrm{LS})}_1$ (fm) & 0.25 & 0.25 \\
    $t^{(\mathrm{LSE})}_1$ (MeV) & $-5101$ & $-11222.2$ \\
    $t^{(\mathrm{LSO})}_1$ (MeV) & $-1897$ & $-4173.4$ \\
    $\mu^{(\mathrm{LS})}_2$ (fm) & 0.40 & 0.40 \\
    $t^{(\mathrm{LSE})}_2$ (MeV) & $-337$ & $-741.4$ \\
    $t^{(\mathrm{LSO})}_2$ (MeV) & $-632$ & $-1390.4$ \\
    \hline
    \multicolumn{3}{l}{\textbf{Tensor force}} \\
    $\mu^{(\mathrm{TN})}_1$ (fm) & 0.40 & 0.40 \\
    $t^{(\mathrm{TNE})}_1$ (MeV\,fm$^{-2}$) & $-1096$ & $-1096$ \\
    $t^{(\mathrm{TNO})}_1$ (MeV\,fm$^{-2}$) & 244 & 244 \\
    $\mu^{(\mathrm{TN})}_2$ (fm) & 0.70 & 0.70 \\
    $t^{(\mathrm{TNE})}_2$ (MeV\,fm$^{-2}$) & $-30.9$ & $-30.9$ \\
    $t^{(\mathrm{TNO})}_2$ (MeV\,fm$^{-2}$) & 15.6 & 15.6 \\
    \hline
    \multicolumn{3}{l}{\textbf{Density-dependent force}} \\
    $\alpha^{(\mathrm{SE})}$ & --- & 1 \\
    $t_\rho^{(\mathrm{SE})}$ (MeV\,fm$^{3}$) & --- & 384\\
    $\alpha^{(\mathrm{TE})}$ & --- & $1/3$ \\
    $t_\rho^{(\mathrm{TE})}$ (MeV\,fm$^{3}$) & --- & 1930 \\
  \end{tabular}
  }
  
\end{table}

The original parameter set in the Gogny-type interaction was defined utilizing the spin- and isospin-exchange operators in Ref. \cite{PhysRevC.21.1568}.
\begin{equation}
  P_{\sigma} = \frac{1+{\bm{\sigma}_1 \cdot \bm{\sigma}_2}}{2}, \quad P_{\tau} = \frac{1+{\bm{\tau}_1 \cdot \bm{\tau}_2}}{2}, \notag
\end{equation}
where $\bm{\sigma}_1$ and $\bm{\sigma}_2$ are the Pauli spin matrices of nucleons 1 and 2, and $\bm{\tau}_1$ and $\bm{\tau}_2$ are the isospin matrices.
We transform the exchange operators into projection operators in Table \ref{tab:gognyparams}.
The projection operators on the singlet-even (SE), triplet-even (TE), singlet-odd (SO), and triplet-odd (TO) channels are given by
\begin{eqnarray}
  P_{\rm{SE}} = \frac{1 - P_{\sigma}}{2} \frac{1 + P_{\tau}}{2}, \quad P_{\rm{TE}} = \frac{1 + P_{\sigma}}{2} \frac{1 - P_{\tau}}{2} \notag \\
  P_{\rm{SO}} = \frac{1 - P_{\sigma}}{2} \frac{1 - P_{\tau}}{2}, \quad P_{\rm{TO}} = \frac{1 + P_{\sigma}}{2} \frac{1 + P_{\tau}}{2}. \notag
\end{eqnarray}

\end{document}